\documentclass[preprint,12pt]{elsarticle}

\usepackage{amsthm} 
\usepackage{amssymb} 
\usepackage{amstext} 
\usepackage{amsmath} 
\usepackage{bm} 
\usepackage{graphicx} 
\usepackage{epstopdf} 
\usepackage{subcaption} 
\usepackage{float}
\usepackage{tikz}
\usepackage{lineno,hyperref}
\usepackage[capitalize]{cleveref} 
\usepackage[makeroom]{cancel} 
\usepackage{multirow} 
\usepackage{booktabs} 
\usepackage{lscape}

\modulolinenumbers[5]

\crefalias{subequation}{equation} 
\crefalias{eqnarray}{equation} 
\crefformat{pluraleq}{Eqs.~(#2#1#3)} 
\Crefformat{pluraleq}{Equations~(#2#1#3)} 

\def\bal#1\nal{\begin{align}#1\end{align}}
\def\bala#1\nala{\begin{align*}#1\end{align*}}
\def\bsub#1\nsub{\begin{subequations}#1\end{subequations}}
\newcommand{\f}{\frac}
\newcommand{\ux}{{\bm x}}
\newcommand{\un}{{\bm n}}
\newcommand{\unab}{{\bf \nabla}}

\newcommand{\uom}{{\bf \Omega}}

\newcommand\ddfrac[2]{\frac{\displaystyle #1}{\displaystyle #2}}

\journal{arXiv}

 \biboptions{sort&compress}

\begin{document}

\begin{frontmatter}

\title{A Spectral Approach for Solving the Nonclassical Transport Equation
}

\author[osu]{R. Vasques\corref{cor1}}
\author[uerj]{L.R.C. Moraes\fnref{moraes}}
\author[uerj]{R.C. Barros\fnref{barros}}
\author[ucb]{R.N. Slaybaugh\fnref{slaybaugh}}

\address[osu]{The Ohio State University, Department of Mechanical and Aerospace Engineering\\ 201 W. 19th Avenue, Columbus, OH 43210}
\address[uerj]{Universidade do Estado do Rio de Janeiro, Departamento de Modelagem Computacional -- IPRJ\\ Rua Bonfim 25, 28625-570, Nova Friburgo, RJ, Brazil}
\address[ucb]{University of California, Berkeley, Department of Nuclear Engineering\\ 4155 Etcheverry Hall, Berkeley, CA 94720-1730}

\cortext[cor1]{Corresponding author: richard.vasques@fulbrightmail.org\\
Postal address: The Ohio State University, Department of Mechanical and Aerospace Engineering, 201 W. 19th Avenue, Columbus, OH 43210}
\fntext[moraes]{lrcmoraes@iprj.uerj.br}
\fntext[barros]{ricardob@iprj.uerj.br}
\fntext[slaybaugh]{slaybaugh@berkeley.edu}

\begin{abstract}

This paper introduces a mathematical approach that allows one to numerically solve the nonclassical transport equation in a deterministic fashion using classical numerical procedures.
The nonclassical transport equation describes particle transport for random statistically homogeneous systems in which the distribution function for free-paths between scattering centers is nonexponential.
We use a spectral method to represent the nonclassical flux as a series of Laguerre polynomials in the free-path variable $s$, resulting in a nonclassical equation that has the form of a classical transport equation. 
We present numerical results that validate the spectral approach, considering transport in slab geometry for both classical and nonclassical problems in the discrete ordinates formulation.
 
\end{abstract}


\end{frontmatter}

\setcounter{section}{0}
\setcounter{equation}{0} 

\section{Introduction}\label{sec1}
\setcounter{equation}{0} 

The theory of \textit{nonclassical} particle transport, which describes processes in which a particle's distance-to-collision is \textit{not} exponentially distributed, has received increased attention in the last decade.
It was originally proposed by Larsen \cite{lar_07} to describe measurements of photon path-length in the Earth's cloudy atmosphere that could not be explained by classical radiative transfer (cf. \cite{davmar_10}).
The theory has been extended over the last few years \cite{larvas_11,fragou_10,vaslar_14a,davxu_14,xudav_16} and has found applications in other areas, including neutron transport in certain types of nuclear reactors \cite{vaslar_09,vas_13,vaslar_14b}, computer graphics \cite{deo_14,jarabo,bitterli}, and problems involving anomalous diffusion (cf. \cite{frasun_16}).
Moreover, a similar kinetic equation has been independently derived for the periodic Lorentz gas in a series of papers by Golse (cf. \cite{gol_12}) and by Marklof and Str\" ombergsson \cite{marstr_10,marstr_11,marstr_14,marstr_15}.

In classical transport, the quantity $dp$ defined as
\bal\label{rev1.1}
dp = \Sigma_t(\ux, E)ds
\nal
represents the incremental probability that a particle located at spatial point $\ux=(x,y,z)$ with energy $E$ will collide while traveling an incremental distance $ds$.
Here, the cross section $\Sigma_t$ is independent of both (i) the particle's direction of flight $\uom = (\Omega_x,\Omega_y,\Omega_z)$, and (ii) the free-path $s$ defined as
\bal\label{rev1.2}
s = \begin{array}{l}
\text{distance traveled by the particle since} \\ \text{its last interaction (birth or scattering).} \end{array}
\nal 
In other words, the instant immediately following the moment in which a particle scatters or is born, its free-path $s$ is set to 0.

The assumption that $\Sigma_t$ is independent of both $\uom$ and $s$ is generally valid when the spatial locations of the scattering centers in the system are uncorrelated (Poisson distributed).
This leads to the Beer-Lambert law, with particle flux decreasing as an exponential function of $s$.
However, a \textit{nonexponential} attenuation law for the particle flux occurs in certain heterogeneous media in which the scattering centers in the system are spatially correlated  \cite{davmar_10,vaslar_09,vas_13,vaslar_14b,deo_14,bitterli,frasun_16,gol_12,
marstr_10,marstr_11,marstr_14,marstr_15}.
The theory of nonclassical particle transport \cite{lar_07,larvas_11,fragou_10,vaslar_14a,davxu_14,xudav_16} was developed to address this class of problems.
In particular, it assumes that $\Sigma_t$ \textit{is} a function of both $\uom$ and $s$, which requires an extended phase space that includes the free-path $s$ as an extra independent variable.

The one-speed nonclassical transport equation with angular-dependent free-paths can be written as \cite{vaslar_14a}
\bsub\label[pluraleq]{1.1}
\bal
\f{\partial }{\partial s}&\Psi(\ux,\uom,s) + \uom\cdot\unab\Psi(\ux,\uom,s) + \Sigma_t(\uom,s)\Psi (\ux,\uom,s) = \label{1.1a}\\
& \delta(s)\left[c\int_{4\pi}\int_0^{\infty}P(\uom'\cdot\uom)\Sigma_t(\uom',s')\Psi(\ux,\uom',s')d\Omega' ds' + \f{Q(\ux)}{4\pi}\right], \quad \ux \in V,\; \uom \in 4\pi, \; 0<s,\nonumber
\nal
where $\Psi$ is the nonclassical angular flux, $c$ is the scattering ratio, and $Q$ is an isotropic source.
Here, $P(\uom'\cdot\uom)d\Omega$ represents the probability that when a particle with direction of flight $\uom'$ scatters, its outgoing direction of flight will lie in $d\Omega$ about $\uom$.
We observe that the Dirac delta function $\delta(s)$ in the right-hand side of \cref{1.1a} follows from the definition given in \cref{rev1.2}; that is, a particle that has just scattered or been born will have its free-path value set to $s=0$.
\Cref{1.1a} is subject to the incident boundary angular flux \cite{larfra_17}
\bal
\Psi(\ux,\uom,s) = \Psi^{b}(\ux,\uom)\delta(s),\quad \ux\in\partial V, \; \un\cdot\uom < 0,\; 0<s.
\nal
\nsub
The angular-dependent nonclassical total cross section $\Sigma_t(\uom,s)$ in \cref{1.1a} satisfies
\bal\label{1.2}
p(\uom,s) = \Sigma_t(\uom,s)e^{-\int_0^s \Sigma_t(\uom,s')ds'},
\nal
where $p(\uom,s)$ is the free-path distribution function in the direction $\uom$.

If \textit{classical} transport takes place, $\Sigma_t$ is independent of both $\uom$ and $s$.
In this case, the free-path distribution reduces to the exponential distribution $p(s) = \Sigma_te^{-\Sigma_ts}$, and \cref{1.1} reduce to the classical linear Boltzmann equation
\bsub\label[pluraleq]{1.3}
\bal
\uom\cdot\unab\Psi_c(\ux,\uom) + \Sigma_t\Psi_c (\ux,\uom) &= c\int_{4\pi}P(\uom'\cdot\uom)\Sigma_t\Psi_c(\ux,\uom')d\Omega' + \f{Q(\ux)}{4\pi},\\
& \hspace{200pt} \ux\in V,\; \uom\in 4\pi,\nonumber\\
\Psi_c(\ux,\uom) &= \Psi^{b}(\ux,\uom),\quad \ux\in\partial V, \;\un\cdot\uom < 0,
\nal
for the classical angular flux
\bal\label{1.3c}
\Psi_c(\ux,\uom) = \int_0^\infty\Psi(\ux,\uom,s)ds.
\nal
\nsub

Numerical results for the nonclassical theory have been provided for diffusion-based approximations and for moment models in the diffusive regime \cite{vaslar_09, larvas_11, vas_13, kryber_13, vaslar_14b, vassla_17a}.
To our knowledge, numerical results for the nonclassical transport equation given by \cref{1.1} are only available for problems in rod geometry \cite{vaskry_15, vassla_16, vaskry_17}.
This is in part due to the difficult task of estimating the nonclassical free-path distribution.
Another reason is that the $s$-dependence of the equation adds to the numerical cost of finding a solution.
For instance, if one interprets the free-path $s$ as a pseudo-time variable, a direct discretization will potentially lead to stiffness since the step size for $s$ will need to be smaller than the spatial step size \cite{vaskry_15}, analogous to the Courant-Friedrichs-Lewy (CFL) condition.
Moreover, the improper integral on the right-hand side of \cref{1.1a} will need to be approximated numerically.

The original contributions of this paper are as follows.
We introduce a mathematical approach that allows one to numerically solve \cref{1.1} in a deterministic fashion using classical numerical procedures.
This approach uses a \textit{spectral} method to represent the nonclassical flux as a series of Laguerre polynomials \cite{hoc_72} in the variable $s$.
The resulting equation has the form of a \textit{classical} transport equation and can be solved by traditional methods. 
Furthermore, it gives one the option to analytically solve the improper integral on the right-hand side of \cref{1.1a}, depending on how hard it is to obtain explicit expressions for the moments of the free-path distribution.
We also present numerical results that validate the spectral approach, considering transport in slab geometry for both classical and nonclassical problems.
To our knowledge, this is the first time deterministic numerical results for  \cref{1.1} are provided for problems in slab geometry.

A summary of the remainder of this paper is given next.
In \cref{sec2}, we present the spectral approach to the nonclassical transport equation.
In \cref{sec3}, we describe the numerical methodology used to solve the transport problems in this paper.
Numerical results are given in \cref{sec4}: in \cref{sec41}, we validate the proposed approach by solving a classical transport problem, and in \cref{sec42} we present numerical results for nonclassical transport problems in a random periodic medium.
We conclude with a discussion in \cref{sec5}.

\section{Spectral Approach}\label{sec2}
\setcounter{equation}{0} 

Let us consider \cref{1.1a} in an equivalent ``initial value'' form \cite{vaslar_14a}:
\bsub\label[pluraleq]{2.1}
\bal
&\f{\partial }{\partial s}\Psi(\ux,\uom,s) + \uom\cdot\unab\Psi(\ux,\uom,s) + \Sigma_t(\uom,s)\Psi (\ux,\uom,s) = 0,\\
& \Psi(\ux,\uom,0)=c\int_{4\pi}\int_0^{\infty}P(\uom'\cdot\uom)\Sigma_t(\uom',s')\Psi(\ux,\uom',s')d\Omega' ds' + \f{Q(\ux)}{4\pi},
\nal
\nsub
and define $\psi$ such that
\bal\label{2.2}
\Psi(\ux,\uom,s)\equiv \psi(\ux,\uom,s)e^{-\int_0^s \Sigma_t(\uom,s')ds'}.
\nal
We can now rewrite the nonclassical problem as
\bsub\label[pluraleq]{2.3}
\bal
&\f{\partial }{\partial s}\psi(\ux,\uom,s) + \uom\cdot\unab\psi(\ux,\uom,s) = 0,\label{2.3a}\\
& \psi(\ux,\uom,0)=  S(\ux,\uom)+\f{Q(\ux)}{4\pi},\label{2.3b}\\
& \psi(\ux,\uom,s) =  \Psi^{b}(\ux,\uom)\delta(s)e^{\int_0^s\Sigma_t(\uom,s')ds'},\quad \ux\in\partial V,\; \un \cdot \uom <0,\label{2.3c}
\nal
where $S(\ux,\uom)$ is the scattering source defined as
\bal
 S(\ux,\uom) = c\int_{4\pi}\int_0^{\infty} P(\uom'\cdot\uom)p(\uom',s')\psi(\ux,\uom',s')d\Omega' ds'
\nal
\nsub
and $p(\uom',s')$ is the free-path distribution given in \cref{1.2}.

To apply the spectral method, we represent $\psi$ as a series of Laguerre polynomials \cite{hoc_72} in $s$:
\bal\label{2.4}
\psi(\ux,\uom,s) = \sum_{m=0}^{\infty} \psi_m(\ux,\uom)L_m(s),
\nal
and replace this ansatz into \cref{2.3}.
The Laguerre polynomials $\{ L_m(s)\}$ are orthogonal with respect to the weight function $e^{-s}$ and satisfy $\f{d}{ds}L_m(s) = \left(\f{d}{ds}-1\right)L_{m-1}(s)$ for $m>0$. 
Therefore, multiplying \cref{2.3a,2.3c} by $e^{-s}L_m(s)$ and operating on them by $\int_0^\infty (\cdot)ds$, we obtain
\bsub\label[pluraleq]{2.5}
\bal
&\uom\cdot\unab\psi_m(\ux,\uom) = \sum_{j=m+1}^\infty \psi_j(\ux,\uom), \quad m=0,1,2,... \,,\\
&\psi_m(\ux,\uom) =  \Psi^{b}(\ux,\uom),\quad \ux\in\partial V,\; \un \cdot \uom <0,\; m=0,1,2,...\,\, .
\nal
\nsub
Moreover, \cref{2.3b} yields
\bal\label{2.6}
\sum_{j=m+1}^\infty\psi_j(\ux,\uom) &= S(\ux,\uom) +\f{Q(\ux)}{4\pi} - \sum_{j=0}^m\psi_j(\ux,\uom), 
\nal
where $S(\ux,\uom) =  c\int_{4\pi}\int_{0}^{\infty}P(\uom'\cdot\uom)p(\uom',s')\textstyle{\sum_{k=0}^{\infty}\limits}\psi_{k}(\ux,\uom')L_{k}(s')ds'd\uom'$.

Substituting \cref{2.6} into \cref{2.5}, we have
\bsub\label[pluraleq]{2.7}
\bal
&\uom\cdot\unab\psi_m(\ux,\uom) + \psi_m(\ux,\uom) = S(\ux,\uom) + \f{Q(\ux)}{4\pi} - \sum_{j=0}^{m-1}\psi_j(\ux,\uom), \quad m=0,1,2,...\,,\\
&\psi_m(\ux,\uom) =  \Psi^{b}(\ux,\uom),\quad \ux\in\partial V,\; \un \cdot \uom <0,\; m=0,1,2,... \,\,.
\nal
\nsub
The nonclassical angular flux $\Psi(\ux,\uom,s)$ is recovered from \cref{2.2,2.4}. 
The classical angular flux $\Psi_c(\ux,\uom)$ is obtained using \cref{1.3c}, such that
\bal\label{2.8}
\Psi_{c}(\ux,\uom) = \int_{0}^{\infty}\Psi(\ux,\uom,s)ds = \int_{0}^{\infty}e^{-\int_{0}^{s}\Sigma_t(\uom,s')ds'}\sum_{m=0}^{\infty}\psi_{m}(\ux,\uom)L_{m}(s)ds\,.
\nal

Rearranging the terms in the scattering source such that
\bal\label{rev2.1}
S(\ux,\uom) = c\int_{4\pi}P(\uom'\cdot\uom)\sum_{k=0}^\infty \psi_k(\ux,\uom')\left[\int_0^\infty p(\uom',s')L_k(s')ds'\right]d\Omega',
\nal
we observe that the improper integral in brackets can be rewritten as a linear combination of the $k$ first raw moments of $p(\uom,s)$.
For example, considering the first three Laguerre polynomials, one obtains
\bsub\label[pluraleq]{rev2.2}
\bal
\int_0^\infty p(\uom',s')L_0(s')ds' &= \int_0^\infty p(\uom',s')ds' = 1\,,\\
\int_0^\infty p(\uom',s')L_1(s')ds' &= \int_0^\infty p(\uom',s')(1-s')ds' = 1 - \tilde{s} (\uom')\,,\\
\int_0^\infty p(\uom',s')L_2(s')ds'&= \int_0^\infty p(\uom',s')\f{1}{2}(s'^2-4s'+2)ds' =\f{\tilde{s^2}(\uom')}{2}  - 2\tilde{s}(\uom') + 1\,,
\nal
\nsub
where $\tilde{s}(\uom)$ and $\tilde{s^2}(\uom)$ are, respectively, the first (mean free-path) and second (mean squared free-path) moments of $p(\uom,s)$ in the direction $\uom$.
In general, it is always preferable to perform the integral in $s$ analytically.
However, $p(\uom,s)$ may be such that obtaining an explicit expression for the moments might be too challenging, or the moments may not even exist.

In order to obtain a numerical solution for \cref{2.7} it is necessary to truncate the series of Laguerre polynomials.
Moreover, in this paper, we have decided to perform the improper integral in \cref{rev2.1} numerically, so that our numerical approach is as problem independent as possible.
The resulting equations can be solved by traditional numerical methods as discussed in the next section.

\section{Numerical Methodology}\label{sec3}
\setcounter{section}{3}
\setcounter{equation}{0}

As the goal of this paper is to provide a first validation of the proposed method, we focus on obtaining numerical solutions to test problems in slab geometry with isotropic scattering and vacuum boundary conditions.
Taking $M$ as the truncation order of the Laguerre expansion, we have:
\begin{subequations}
	\begin{align}
&	\mu\frac{\partial}{\partial x}\psi_{m}(x,\mu) + \psi_{m}(x,\mu) =  S(x) + \f{Q(x)}{2} - \sum_{j=0}^{m-1}\psi_{j}(x,\mu),\, m = 0,1,\dots M,\label{3.1a} \\
&	\psi_{m}(-X,\mu)  = 0, \mu>0, \, m = 0,1,\dots,M,\label{3.1b} \\
&	\psi_{m}(X,\mu)  = 0, \mu<0, \, m = 0,1,\dots,M,\label{3.1.c} \\
&S(x) =  \frac{c}{2}\int_{-1}^{1}\int_{0}^{\infty}\Sigma_{t}(\mu^{\prime},s^{\prime})e^{-\int_{0}^{s^{\prime}}\Sigma_{t}(\mu^{\prime},s'')ds''}\sum_{k=0}^{M}   \psi_{k}(x,\mu^{\prime})L_{k}(s^{\prime})ds^{\prime}d\mu^{\prime},\label{3.1d}\\
&\Psi_{c}(x,\mu) = \int_{0}^{\infty}e^{-\int_{0}^{s}\Sigma_{t}(\mathbf{\mu},s^{\prime})ds^{\prime}}\sum_{m=0}^{M}\psi_{m}(x,\mathbf{\mu})L_{m}(s)ds.\label{3.1e}
\end{align}
\label[pluraleq]{3.1}
\end{subequations}
Next, we write \cref{3.1} in the conventional discrete ordinates (S$_{N}$) formulation \cite{lewis}:
\begin{subequations}
	\begin{align}
	&	\mu_{l}\frac{d}{d x}\psi_{m,l}(x) + \psi_{m,l}(x) =  S(x) + \f{Q(x)}{2} - \sum_{j=0}^{m-1}\psi_{j,l}(x),\, m = 0,1,\dots M,\: l = 1,2,\dots N,\label{3.2a} \\
	&	\psi_{m,l}(-X)  = 0,  \, m = 0,1,\dots,M,\: l = 1,2,\dots \frac{N}{2},\label{3.2b} \\
	&	\psi_{m,l}(X)  = 0,  \, m = 0,1,\dots,M,\: l = \frac{N}{2}+1,\dots N, \label{3.2c}\\
	&S(x) = \frac{c}{2}\sum_{n=1}^{N}\omega_{n}\sum_{k=0}^{M}\psi_{k,n}(x)\int_{0}^{\infty}\Sigma_{t_{n}}(s^{\prime})L_{k}(s^{\prime})e^{-\int_{0}^{s^{\prime}}\Sigma_{t_{n}}(s'')ds''}ds^{\prime},\label{3.2d}\\
	&\Psi_{c,l}(x) =\sum_{m=0}^{M}\psi_{m,l}(x)\int_{0}^{\infty}L_{m}(s)e^{-\int_{0}^{s}\Sigma_{t_{l}}(s^{\prime})ds^{\prime}}ds.\label{3.2e}
	\end{align}
	\label[pluraleq]{3.2}
\end{subequations}
Here, we have defined $\psi_{m,l}(x)$ as $\psi_{m}(x,\mu_{l})$, where the direction-of-motion variable $\mu$ has been discretized in $N$ discrete values $\mu_{l}$. 
Similarly, $\Sigma_{t_{l}}(s^{\prime})$ has been defined as $\Sigma_{t}(\mu_{l},s^{\prime})$, and the angular integral in \cref{3.1d} has been approximated by the angular quadrature formula with weights $\omega_{n}$ in \cref{3.2d}.

Furthermore, we take two steps to approximate the improper integrals in \cref{3.2d,3.2e}: (i) we truncate the upper limit to a finite number $L$ by neglecting the integral in the complementary range $(L,\infty)$; and (ii) we perform the simple linear transformation $z = \frac{2}{L}s^{\prime}-1$ to translate the interval $[0,L]$ into $[-1,1]$.
We use  the Gauss-Legendre quadrature \cite{burden} in \cref{3.2d} to write
\begin{equation}
\begin{gathered}
\frac{L}{2}\int_{-1}^{1}\Sigma_{t_{n}}\left(\left(z+1\right)\frac{L}{2}\right)L_{k}\left(\left(z+1\right)\frac{L}{2}\right)e^{-\int_{0}^{\left(z+1\right)\frac{L}{2}}\Sigma_{t_{n}}(s'')ds''}dz = \\
\frac{L}{2}\sum_{i=1}^{G_{\ell}}\Sigma_{t_{n}}\left(\left(z_{i}+1\right)\frac{L}{2}\right)L_{k}\left(\left(z_{i}+1\right)\frac{L}{2}\right)e^{-\int_{0}^{\left(z_{i}+1\right)\frac{L}{2}}\Sigma_{t_{n}}(s'')ds''}\omega_{i},
\end{gathered}
\label{3.3}
\end{equation}
where $G_{\ell}$ is the order of the Gauss-Legendre quadrature.

A similar procedure is carried out for the improper integral in \cref{3.2e}.
\Cref{3.2d,3.2e} appear as
\begin{equation}
S(x) = \frac{cL}{4}\sum_{n=1}^{N}\omega_{n}\sum_{k=0}^{M}\psi_{k,n}(x)\sum_{i=1}^{G_{\ell}}\Sigma_{t_{n}}\left(\left(z_{i}+1\right)\frac{L}{2}\right)L_{k}\left(\left(z_{i}+1\right)\frac{L}{2}\right)e^{-\int_{0}^{\left(z_{i}+1\right)\frac{L}{2}}\Sigma_{t_{n}}(s'')ds''}\omega_{i}\label{3.4}
\end{equation}
and
\begin{equation}
\psi_{c,l}(x) = \frac{L}{2}\sum_{m=0}^{M}\psi_{m,l}(x)\sum_{i=1}^{G_{\ell}}L_{k}\left(\left(z_{i}+1\right)\frac{L}{2}\right)e^{-\int_{0}^{\left(z_{i}+1\right)\frac{L}{2}}\Sigma_{t_{n}}(s'')ds''}\omega_{i}.\label{3.5}
\end{equation}
\Cref{3.4} is used in the S$_{N}$ equations (3.2a), which are solved numerically using the conventional fine-mesh diamond difference (DD) method with the source iteration (SI) scheme \cite{lewis}.
In addition, the classical angular flux in direction $\mu_{l}$ is approximated by \cref{3.5}.

At this point we consider a spatial discretization grid on the slab, wherein each cell has width $h_{j}$, $j=1,2,\dots J$.
Applying the operator $\frac{1}{h_{j}}\int_{x_{j-1/2}}^{x_{j+1/2}} (\cdot) dx$, with $h_{j} = x_{j+1/2}-x_{j-1/2}$, to the S$_{N}$ equations (3.2a), we obtain the conventional discretized spatial balance S$_{N}$ equations, where we define the cell-average quantity
\begin{equation}
\bar\psi_{m,l,j} = \frac{1}{h_{j}}\int_{x_{j-1/2}}^{x_{j+1/2}}\psi_{m,l}(x)dx. \label{3.6}
\end{equation}
As with the DD method, we approximate $\psi_{m,l}(x)$ by piecewise continuous linear functions across the spatial grid.
In other words, we consider the approximation 
\begin{gather}
\bar\psi_{m,l,j} = \ddfrac{\psi_{m,l,j+1/2}+\psi_{m,l,j-1/2}}{2}. \label{3.7}
\end{gather}
By substituting \cref{3.7} into the discretized balance equations, we obtain the nonclassical S$_{N}$ sweep equations 
\begin{subequations}
	\begin{align}		
\psi_{m,l,j+1/2} = \ddfrac{\bar{S}_{j} + Q_{j} - \sum_{k=0}^{m-1}\bar{\psi}_{k,l,j} - \psi_{m,l,i-1/2}\left(\frac{1}{2}-\frac{\mu_{l}}{h_{j}}\right)}{\left(\frac{1}{2}+\frac{\mu_{l}}{h_{j}}\right)},\label{3.8a}\\ m = 0:M,\, l=1:\frac{N}{2}, \,j=1:J,\nonumber\qquad\:\\
	\psi_{m,l,j-1/2} = \ddfrac{\bar{S}_{j} + Q_{j} - \sum_{k=0}^{m-1}\bar{\psi}_{k,l,j} - \psi_{m,l,j+1/2}\left(\frac{1}{2}-\frac{|\mu_{l}|}{h_{j}}\right)}{\left(\frac{1}{2}+\frac{|\mu_{l}|}{h_{j}}\right)} \label{3.8b},\\  m = 0:M,\, l=\frac{N}{2}+1:N, \,j=J:1,\nonumber\:\:\:\,
	\end{align}
		\label[pluraleq]{3.8}
\end{subequations}
with
\begin{equation}
\begin{gathered}
\bar{S}_{j} = \frac{c}{2}\sum_{n=1}^{N}\omega_{n}\sum_{k=0}^{M}\bar{\psi}_{k,n,j} \sum_{i=1}^{G_{\ell}}\Sigma_{t_{n}}\left(\left(z_{i}+1\right)\frac{L}{2}\right)L_{k}\left(\left(z_{i}+1\right)\frac{L}{2}\right)e^{-\int_{0}^{\left(z_{i}+1\right)\frac{L}{2}}\Sigma_{t_{n}}(s'')ds''}\omega_{i}, \\\,j=1:J. \label{3.9}
\end{gathered}
\end{equation}

We use \cref{3.8a} to sweep from left to right $\left(\mu_{l}>0\right)$ and \cref{3.8b} to sweep from right to left $\left(\mu_{l}<0\right)$, updating $\bar{S}_{j},\,j=1:J$, and the summation terms, until a prescribed stopping criterion is satisfied.
The stopping criterion adopted is that the relative deviations between two consecutive estimates of classical scalar fluxes in each point of the spatial discretization grid need to be smaller than or equal to a prescribed positive constant $\xi$.

\section{Numerical Results}\label{sec4}
\setcounter{section}{4}
\setcounter{equation}{0}

In this section we provide numerical results that validate the spectral approach and the proposed numerical methodology.
First, we apply this approach to solve a classical transport problem in a homogeneous slab and show that it correctly estimates the scalar flux. 
Then, we proceed to solve a nonclassical transport problem in a one-dimensional, random periodic slab.
For all numerical experiments in this section we adopt $\xi = 1\times 10^{-6}$ for the stopping criterion, $N = 20$, $G_{\ell}=300$, and truncate the $s$ range in the length $L$ of the slab.
The spatial domain is discretized in each problem to yield $h_{j} = 0.005$ cm, $j=1,2,...,J$.

\subsection{Validation: Classical Transport}\label{sec41}

\begin{figure}[htb]
	\centering
	\includegraphics[scale=0.7]{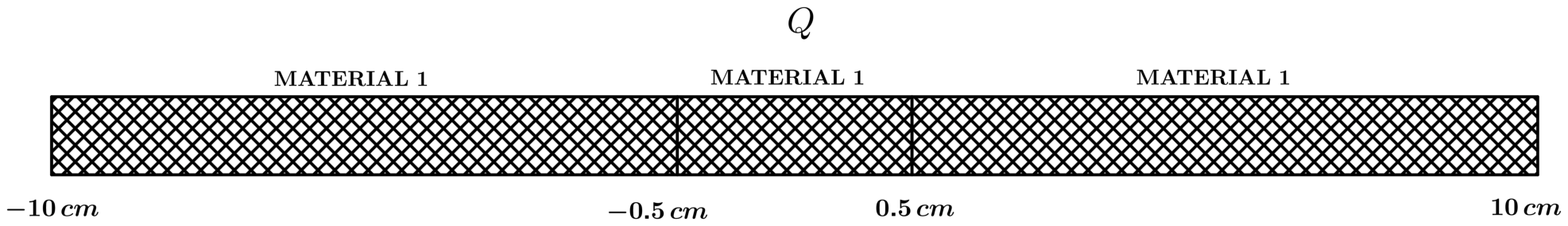}
		\caption{One-dimensional classical slab.} 
	\label{fig1}
\end{figure}

In order to validate the proposed methodology, we apply the numerical approach described in the previous section to solve a \textit{classical} transport problem in slab geometry.
The slab is composed of a homogeneous material (material 1), with total cross section $\Sigma_{t_{1}} = 1$ cm$^{-1}$, as depicted in \cref{fig1}.
A neutron source $Q(x) = 1\times10^{17}$ neutrons/cm$^{3}\cdot$s , with $-0.5$ cm $< x < 0.5$ cm, is located in a region at the center of the slab. 
We emphasize that, as $\Sigma_{t_{1}}$ is independent of $\mu$ and $s$ in all spatial domain, the free-path distribution $p(\mu,s)$ reduces to the exponential distribution $p(s) = \Sigma_{t_{1}}e^{-\Sigma_{t_{1}}s}$.
 
We are interested in how accurately the nonclassical model predicts the classical scalar flux.
To this end, we compare the results obtained with the solution of the classical linear Boltzmann equation in slab geometry.
\Cref{tab1} displays the classical scalar fluxes for two different values of the scattering ratio $c$ and three different values of the truncation order of the Laguerre expansion $M$.
One can see that, even for low values of $M$, the results generated by the nonclassical S$_{20}$ transport equations using the proposed methodology are in close agreement with the results obtained by solving the classical S$_{20}$ transport equations.
This is expected and provides a first validation of the nonclassical spectral approach, since this choice of parameters means the nonclassical transport equation should reduce to the classical linear Boltzmann equation.

\subsection{Nonclassical Transport in a Random Periodic Slab}\label{sec42}

\begin{figure}[htb]
	\centering
	\includegraphics[scale=0.7]{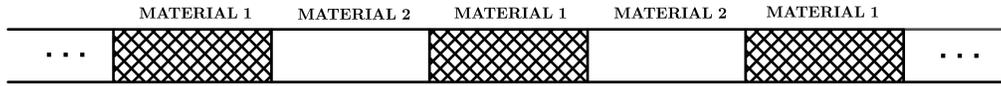}
	\caption{One-dimensional Random Periodic Media.} 
	\label{fig2}
\end{figure}

Let us consider a one-dimensional physical system similar to the
one introduced in \cite{zuchuat}, composed of two distinct materials periodically arranged.
The period is given by $\ell = \ell_{1}+\ell_{2}$, with $\ell_{1}$ and $\ell_{2}$ representing the width of each material.
Material 1 is a solid with $\Sigma_{t_{1}} = 1$ cm$^{-1}$, and material 2 is defined as void, i.e., $\Sigma_{t_{2}} = 0$ cm$^{-1}$.
This periodic system is \textit{randomly placed} in the infinite line $-\infty<x<\infty$, such that the probability $P_{i}$ of finding material $i\in \{1,2\}$ in a given point is $\ell_{i}/\ell$.
Therefore, material parameters (such as the cross sections) are stochastic functions of space.
\Cref{fig2} illustrates this periodic system. 

The ensemble-averaged free-path distribution for the problem depicted in \cref{fig2} has been analytically calculated in \cite{vaskry_17} to different material widths, with expressions for $p(\mu,s)$ given by
\begin{subequations}\label[pluraleq]{4.1}
\begin{itemize}
\item Case 1: $\ell_1<\ell_2$
\end{itemize}
\begin{align}
p(\mu,s) = \left\{
\begin{array}{ll}
\frac{\Sigma_{t1}}{\ell_1}(n\ell +\ell_1-s|\mu|)e^{-\Sigma_{t1}(s-n\ell_2/|\mu|)}, & \text{if } n\ell\leq s|\mu| \leq n\ell+\ell_1\\
0, & \text{if } n\ell+\ell_1 \leq s|\mu| \leq n\ell+\ell_2\\
\frac{\Sigma_{t1}}{\ell_1}(s|\mu|-n\ell-\ell_2)e^{-\Sigma_{t1}[s-(n+1)\ell_2/|\mu|]}, & \text{if } n\ell+\ell_2 \leq s|\mu| \leq (n+1)\ell\\
\end{array}
\right.
\end{align}
\begin{itemize}
\item Case 2: $\ell_1=\ell_2$
\end{itemize}
\begin{align}
p(\mu,s) = \left\{
\begin{array}{ll}
\frac{\Sigma_{t1}}{\ell_1}(n\ell +\ell_1-s|\mu|)e^{-\Sigma_{t1}(s-n\ell_2/|\mu|)}, & \text{if } n\ell\leq s|\mu| \leq n\ell+\ell_1\\
\frac{\Sigma_{t1}}{\ell_1}(s|\mu|-n\ell-\ell_2)e^{-\Sigma_{t1}[s-(n+1)\ell_2/|\mu|]}, & \text{if } n\ell+\ell_2 \leq s|\mu| \leq (n+1)\ell\\
\end{array}
\right.
\end{align}
\begin{itemize}
\item Case 3: $\ell_1>\ell_2$
\end{itemize}
\begin{align}
p(\mu,s) = \left\{
\begin{array}{ll}
\frac{\Sigma_{t1}}{\ell_1}(n\ell +\ell_1-s|\mu|)e^{-\Sigma_{t1}(s-n\ell_2/|\mu|)}, & \\
\hspace{6cm}\text{if } n\ell\leq s|\mu| \leq n\ell+\ell_2 & \\
\frac{\Sigma_{t1}}{\ell_1}[(n\ell +\ell_2-s|\mu|)(1-e^{\Sigma_{t1}\ell_2/|\mu|}) +\ell_1-\ell_2]e^{-\Sigma_{t1}(s-n\ell_2/|\mu|)}, & \\
\hspace{6cm}\text{if } n\ell+\ell_2\leq s|\mu| \leq n\ell+\ell_1 & \\
\frac{\Sigma_{t1}}{\ell_1}(s|\mu|-n\ell-\ell_2)e^{-\Sigma_{t1}[s-(n+1)\ell_2/|\mu|]}, & \\
\hspace{6cm}\text{if } n\ell+\ell_1 \leq s|\mu| \leq (n+1)\ell & 
\end{array}
\right.
\end{align}
\end{subequations}
where $n=0, 1, 2, ...$ .

We perform numerical experiments considering the two sets of problems (A and B) displayed in \cref{tab2}.
In each problem, we consider the existence of a neutron source defined as \cite{vaskry_17}
\begin{gather}
Q(x) = \left\{\begin{array}{l}
\frac{\ell_{1}}{\ell}\times 10^{17}\, \text{neutrons/cm}^{3}\cdot\text{s},\, \text{if} \, x_{1} \leq x \leq x_{2}; \\
0,\,\text{otherwise};
\end{array}\right., 
\end{gather}
where $x_{1}$ and $x_{2}$ are spatial points of the domain.

Here, we are interested in how accurately the nonclassical model predicts the ensemble-averaged scalar flux over all physical realizations of the random medium.
To this end, we compare the nonclassical results against benchmark results obtained by averaging the solutions of the classical transport equation over a large number of physical realizations of this random system.
The benchmarks were produced by solving $1/h_{j}$ classical transport problems.
Details of how to obtain the benchmark solution can be found in \cite{vaskry_17}.

\begin{figure}
\begin{subfigure}{.48\textwidth}
		\centering
	\includegraphics[width=1\linewidth,height=0.26\textheight]{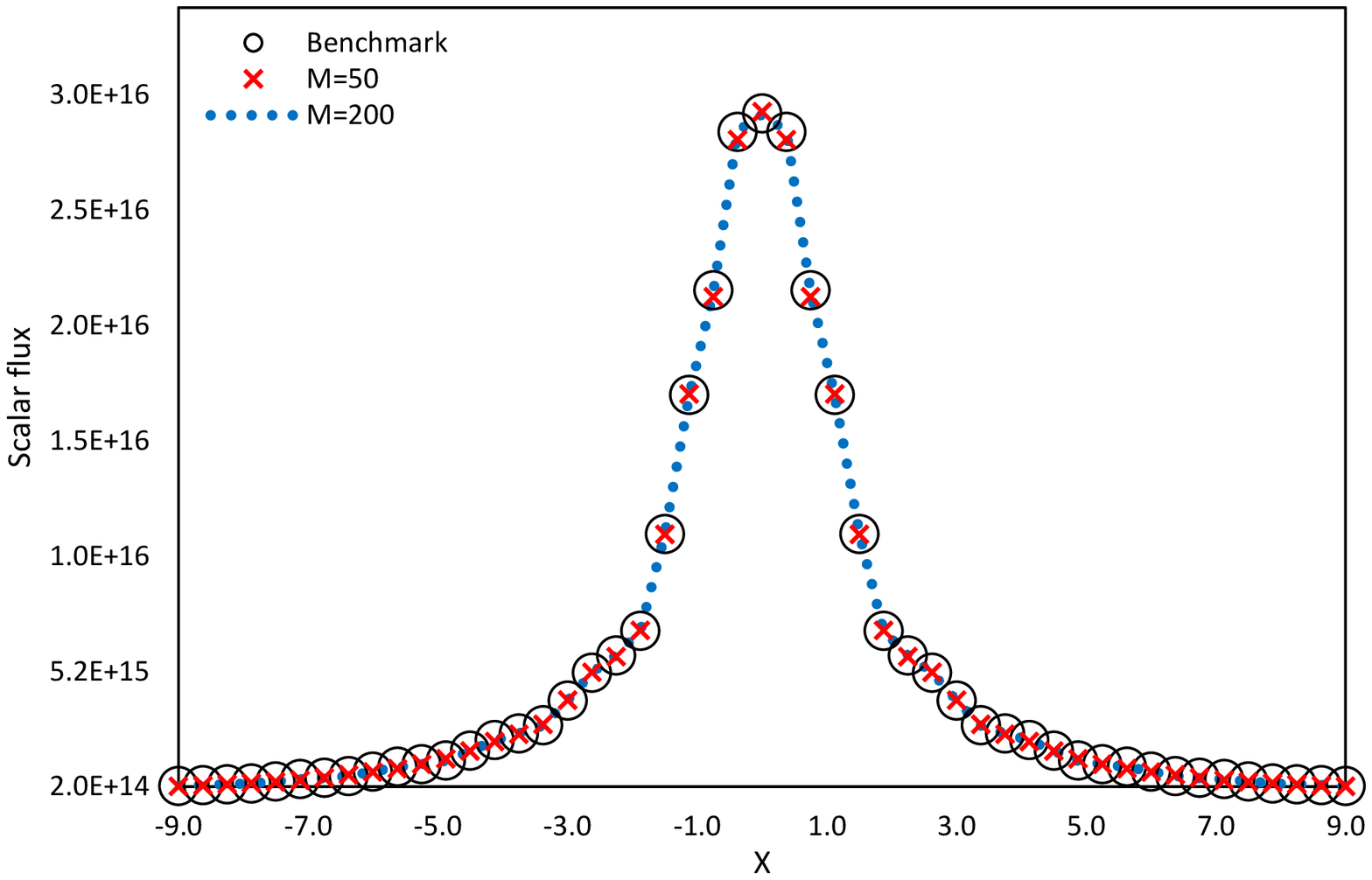}
	\caption{EASF for problem A$_{1}$ with $c = 0.0$.}
\end{subfigure}
\begin{subfigure}{.48\textwidth}
	\centering
	\includegraphics[width=1\linewidth,height=0.26\textheight]{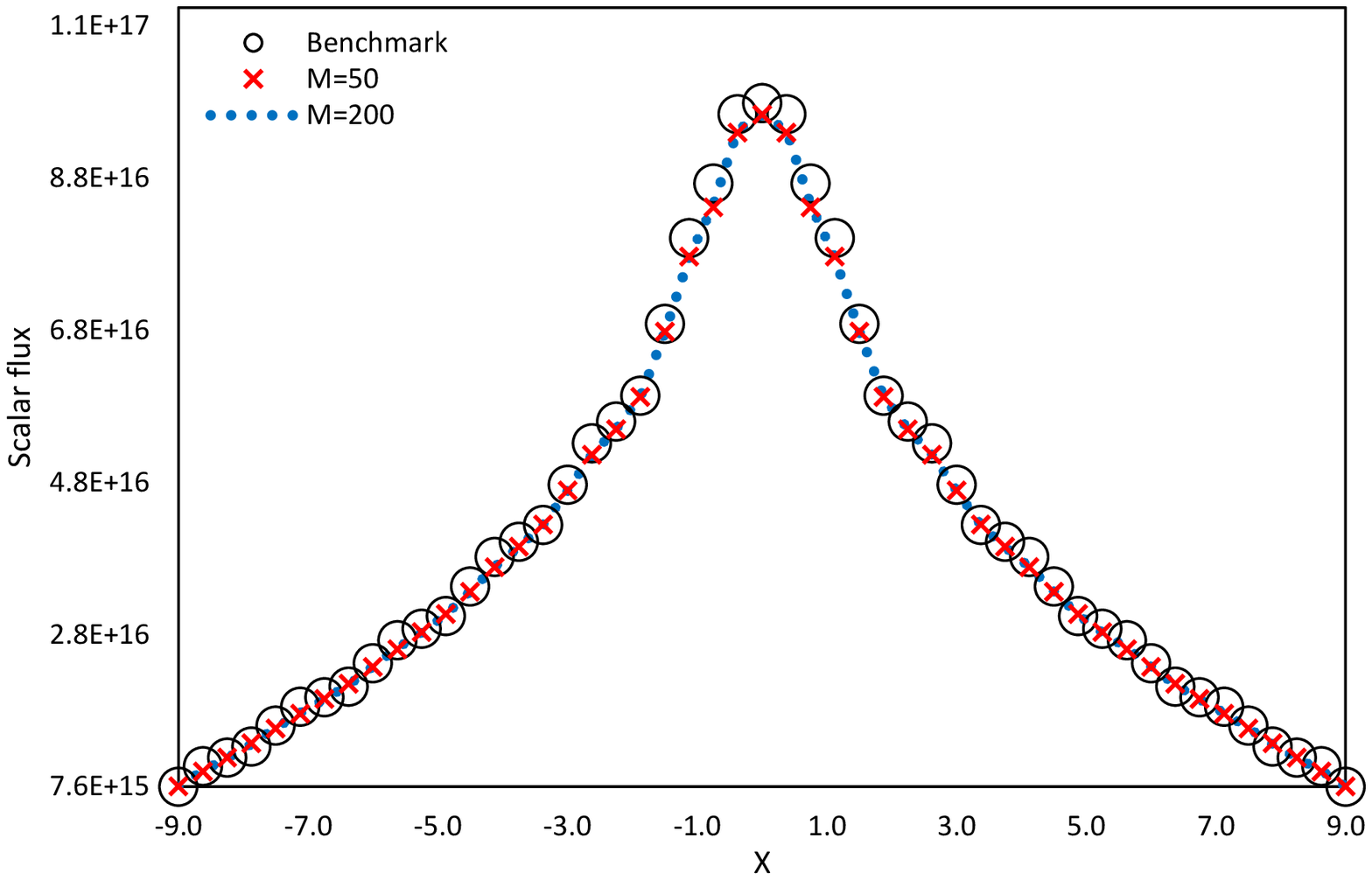}
	\caption{EASF for problem A$_{1}$ with $c = 0.9$.}
\end{subfigure}

\begin{subfigure}{.48\textwidth}
	\centering
	\includegraphics[width=1\linewidth,height=0.26\textheight]{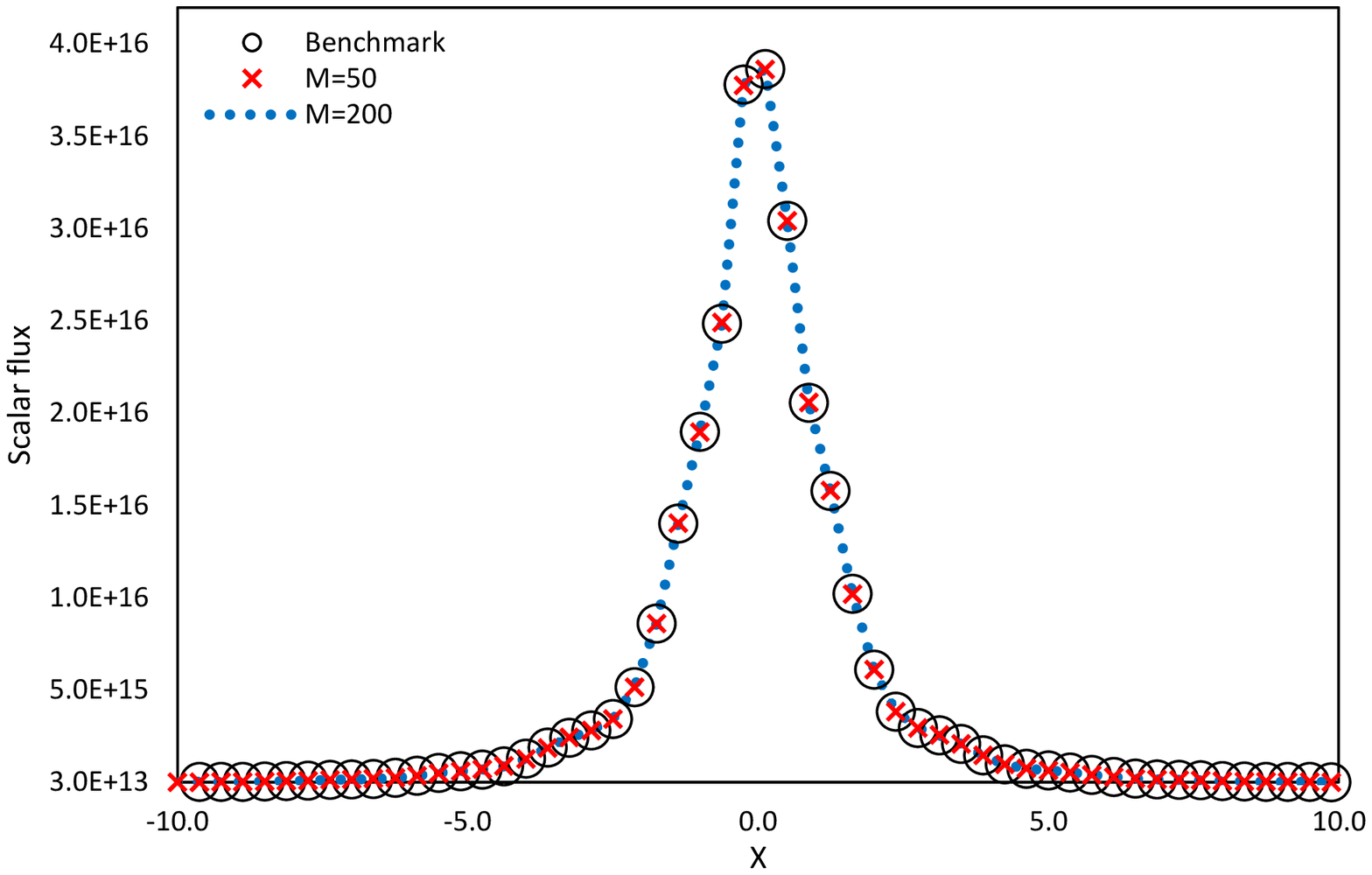}
	\caption{EASF for problem A$_{2}$ with $c = 0.0$.}
\end{subfigure}
\begin{subfigure}{.48\textwidth}
	\centering
	\includegraphics[width=1\linewidth,height=0.26\textheight]{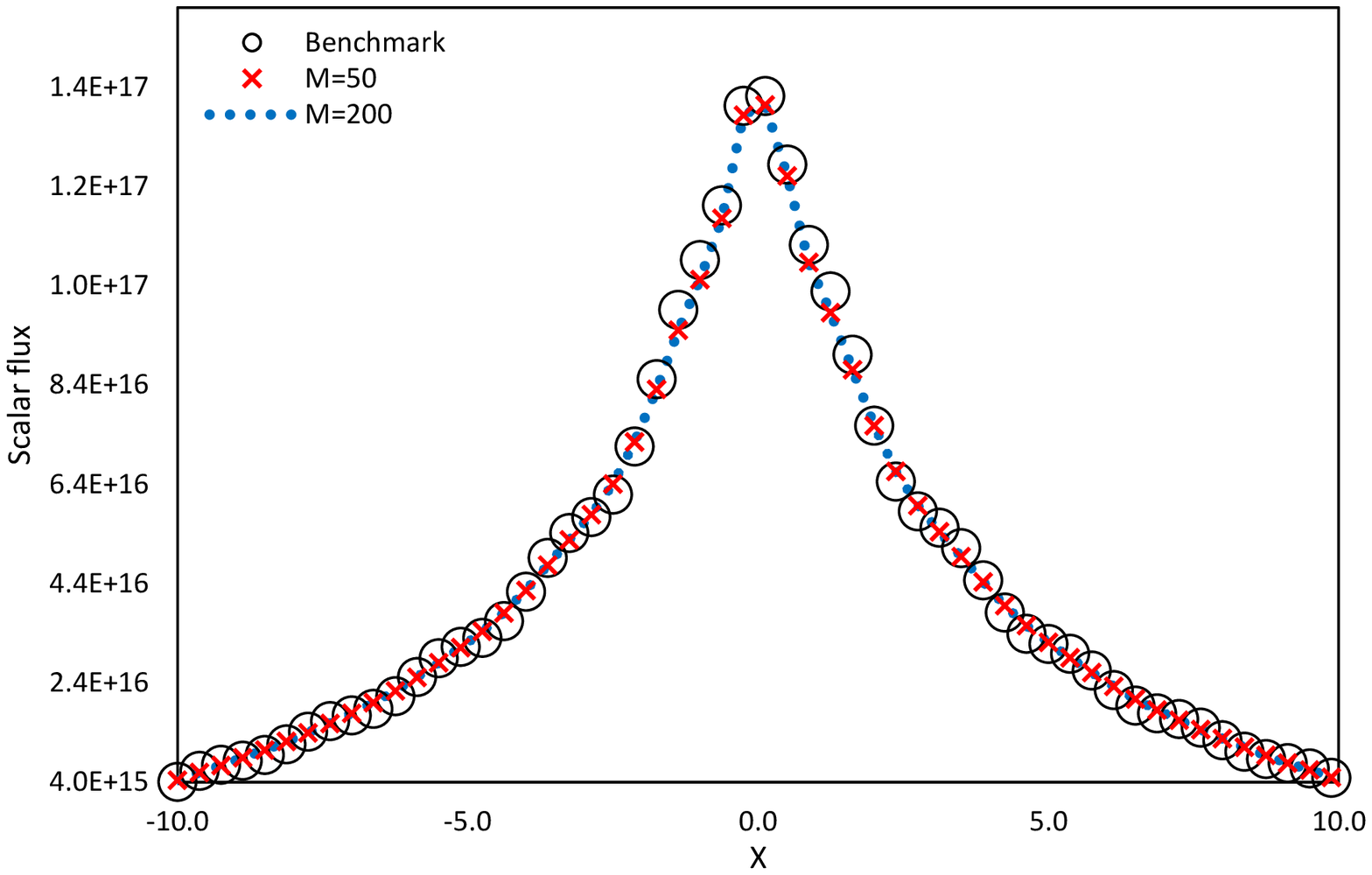}
	\caption{EASF for problem A$_{2}$ with $c = 0.9$.}
\end{subfigure}

\begin{subfigure}{.48\textwidth}
	\centering
	\includegraphics[width=1\linewidth,height=0.26\textheight]{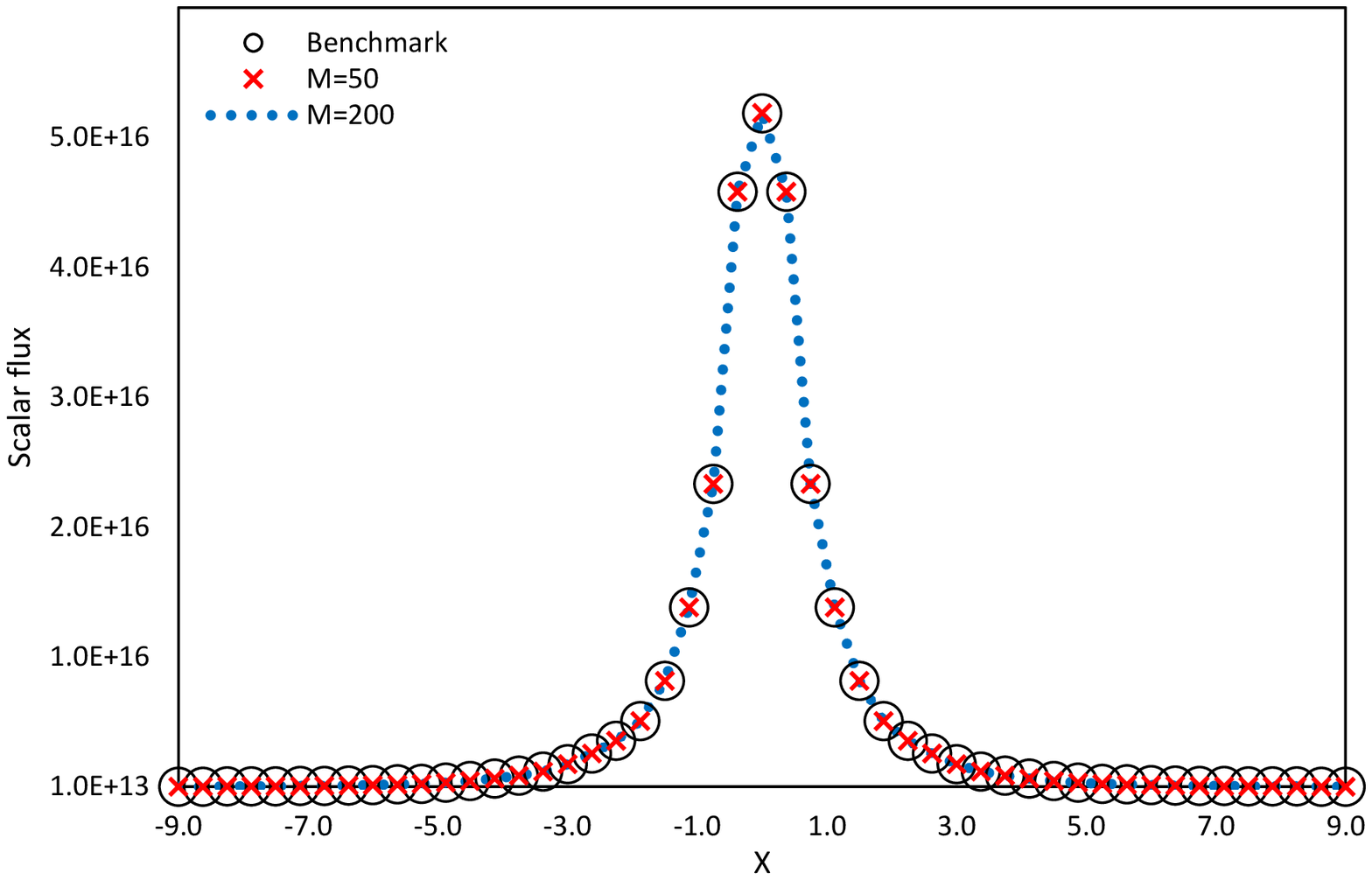}
	\caption{EASF for problem A$_{3}$ with $c = 0.0$.}
\end{subfigure}
\begin{subfigure}{.48\textwidth}
	\centering
	\includegraphics[width=1\linewidth,height=0.26\textheight]{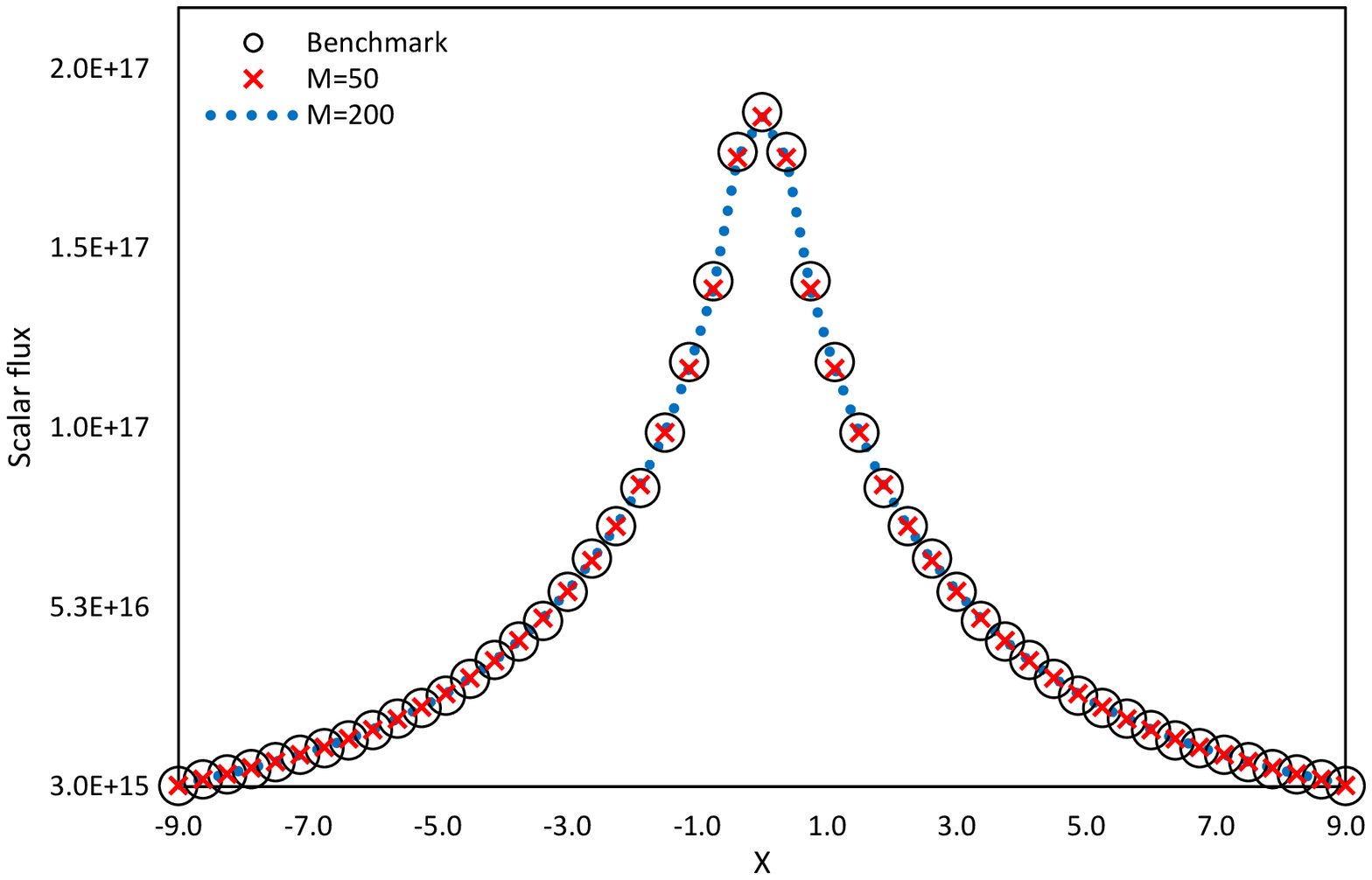}
	\caption{EASF for problem A$_{3}$ with $c = 0.9$.}
\end{subfigure}

\caption{Ensemble-averaged scalar fluxes (EASF) for problem set A.
}\label{fig3}
\end{figure}

\begin{figure}
	\begin{subfigure}{.48\textwidth}
		\centering
		\includegraphics[width=1\linewidth,height=0.26\textheight]{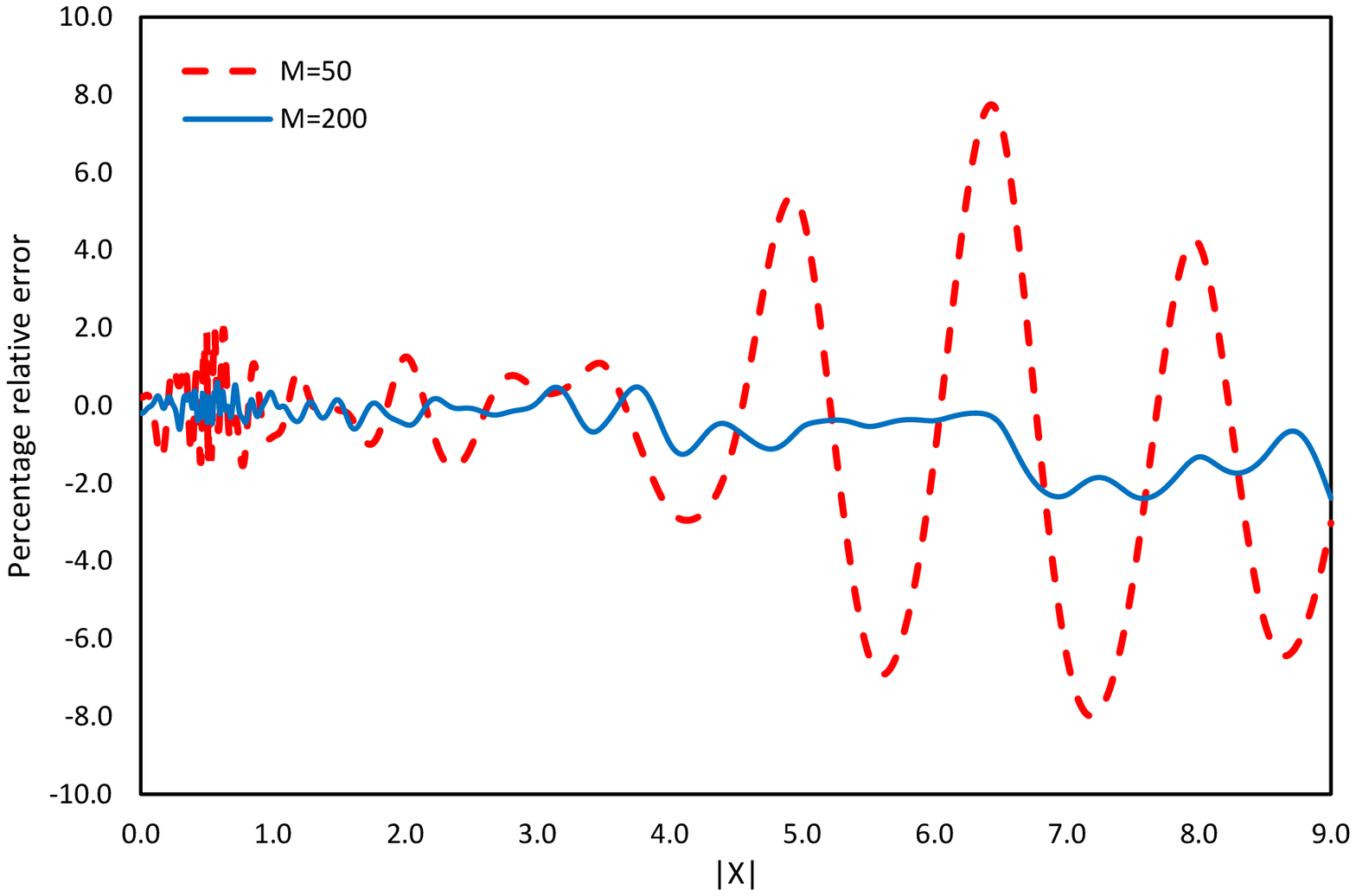}
		\caption{PRE for problem A$_{1}$ with $c = 0.0$.}
	\end{subfigure}
	\begin{subfigure}{.48\textwidth}
		\centering
		\includegraphics[width=1\linewidth,height=0.26\textheight]{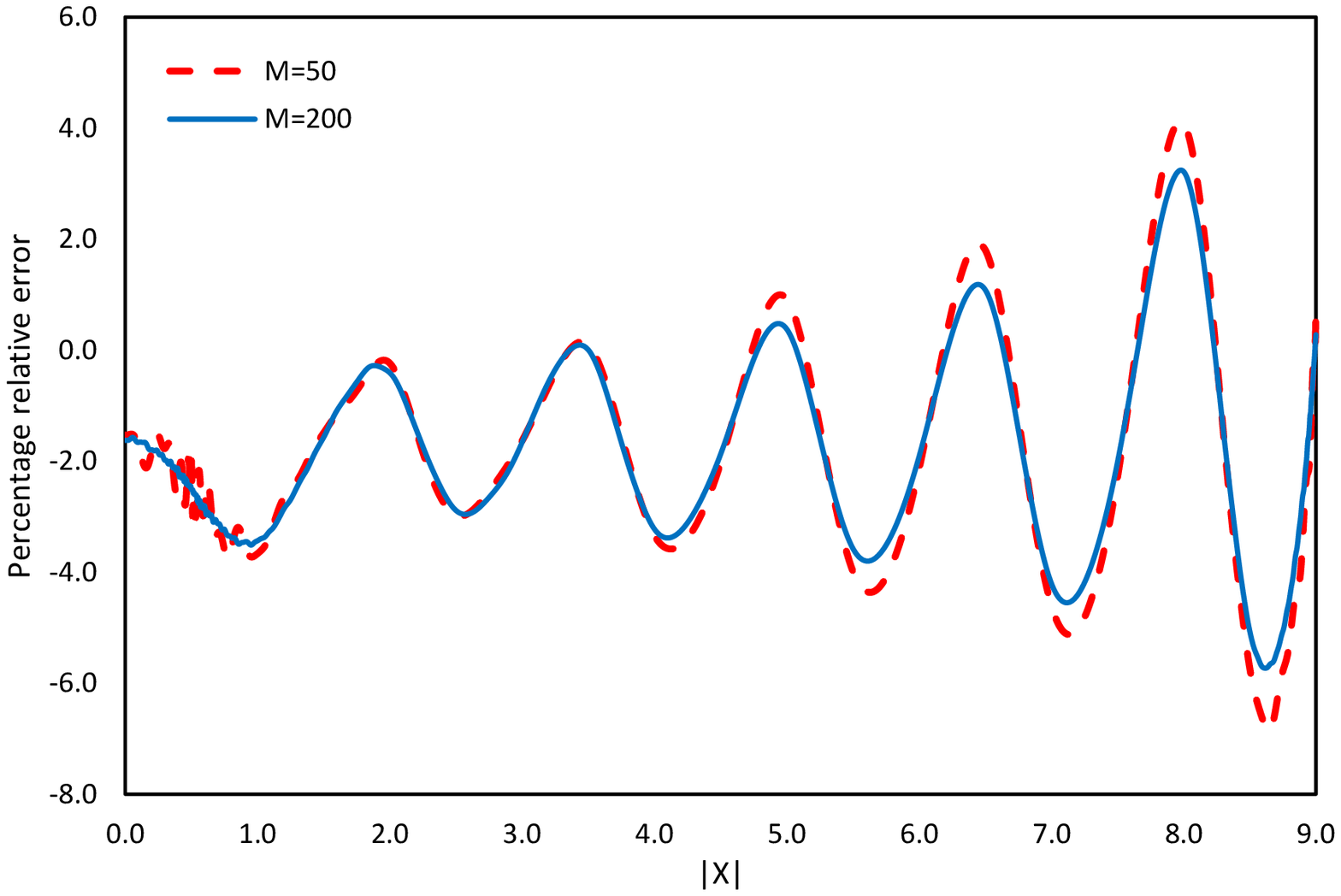}
		\caption{PRE for problem A$_{1}$ with $c = 0.9$.}
	\end{subfigure}
	
	\begin{subfigure}{.48\textwidth}
		\centering
		\includegraphics[width=1\linewidth,height=0.26\textheight]{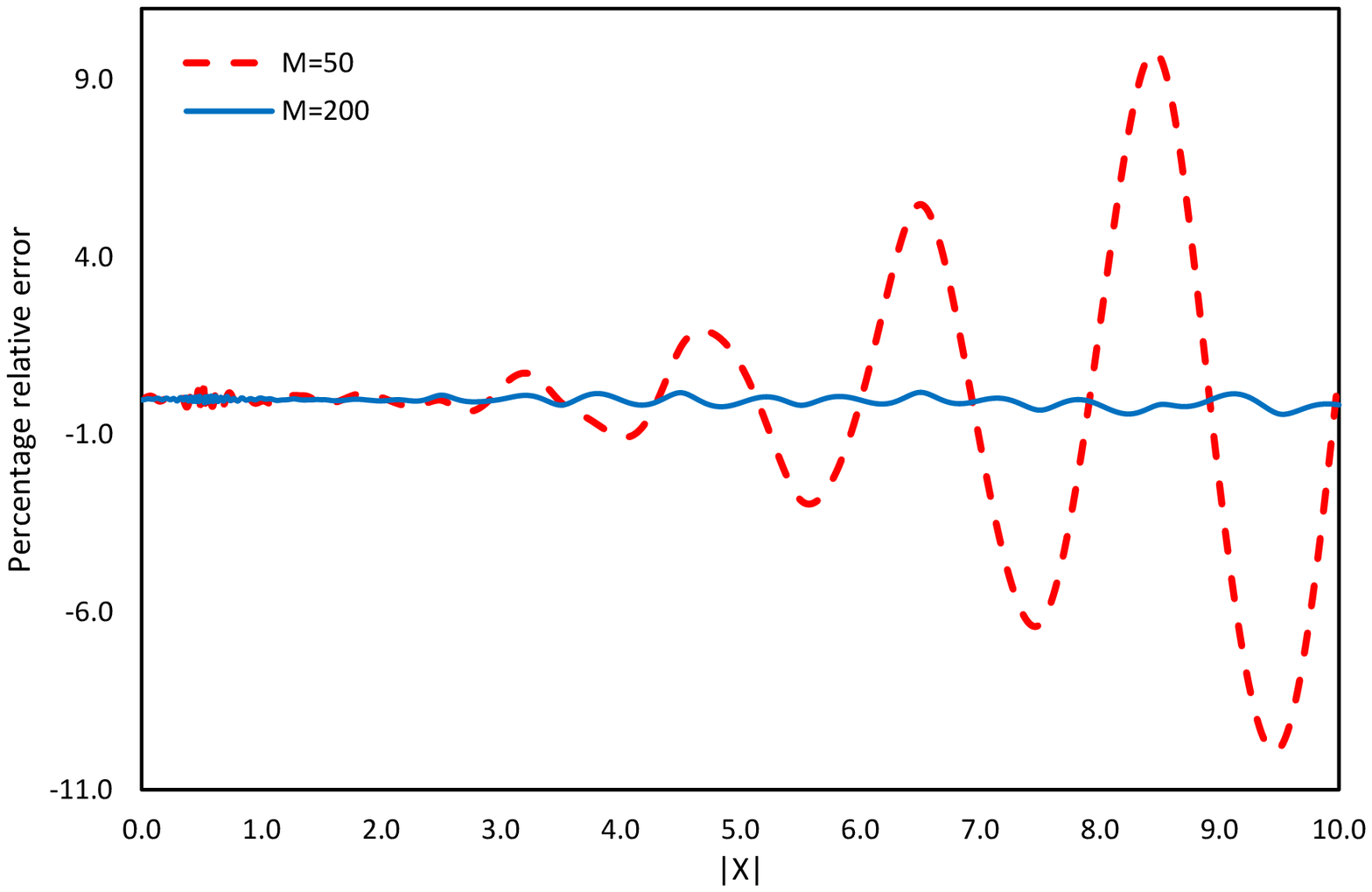}
		\caption{PRE for problem A$_{2}$ with $c = 0.0$.}
	\end{subfigure}
	\begin{subfigure}{.48\textwidth}
		\centering
		\includegraphics[width=1\linewidth,height=0.26\textheight]{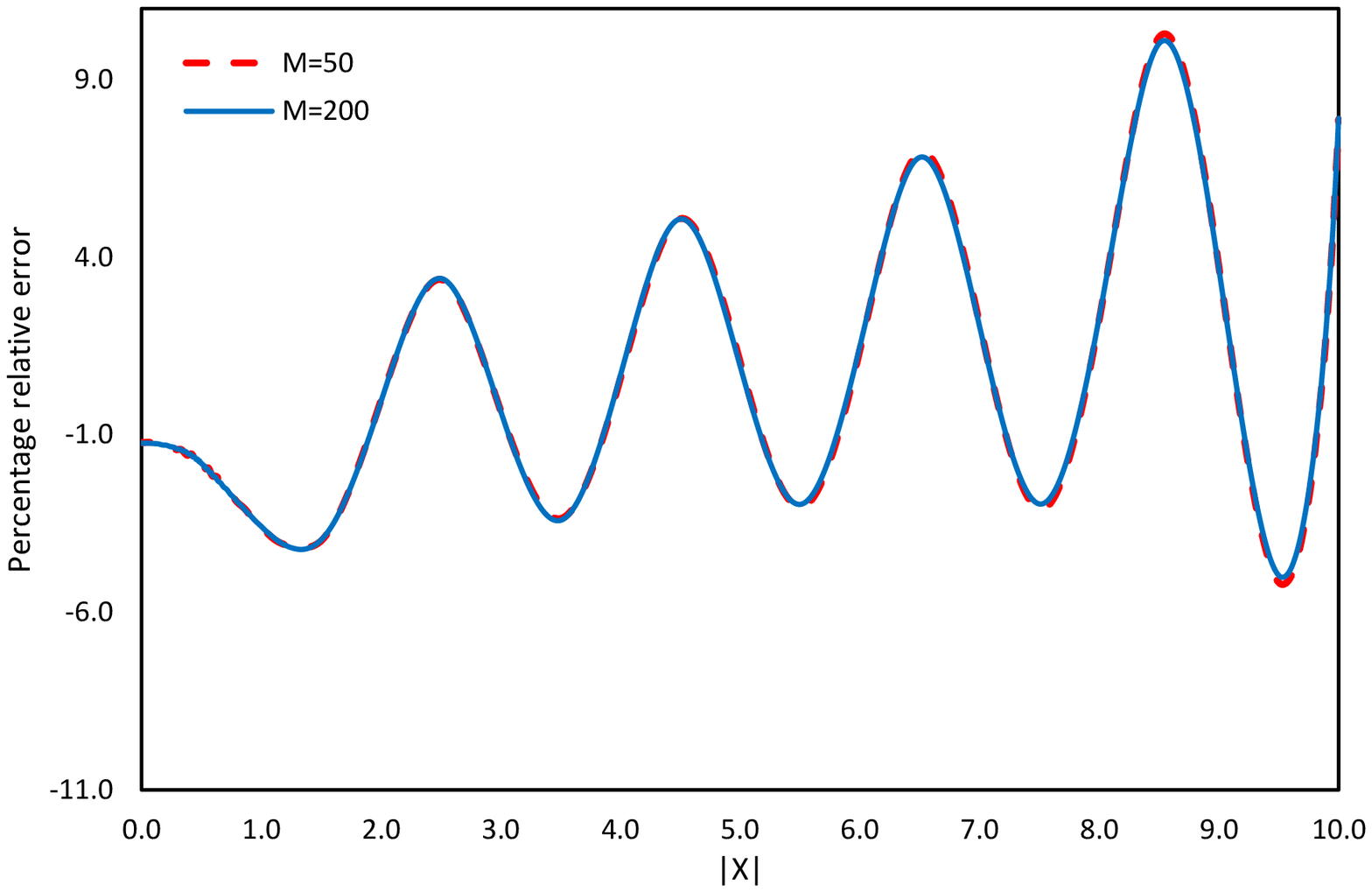}
		\caption{PRE for problem A$_{2}$ with $c = 0.9$.}
	\end{subfigure}
	
	\begin{subfigure}{.48\textwidth}
		\centering
		\includegraphics[width=1\linewidth,height=0.26\textheight]{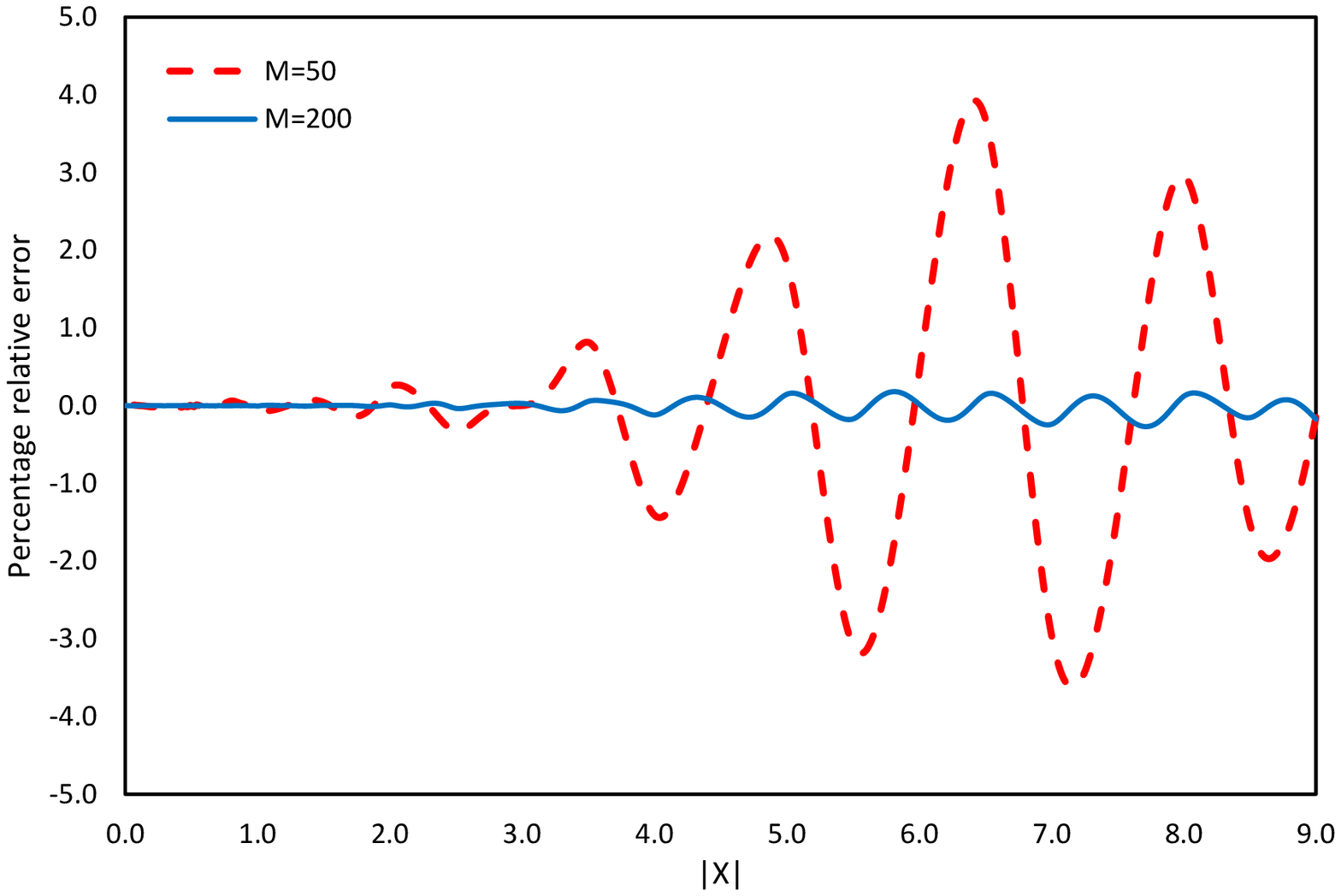}
		\caption{PRE for problem A$_{3}$ with $c = 0.0$.}
	\end{subfigure}
	\begin{subfigure}{.48\textwidth}
		\centering
		\includegraphics[width=1\linewidth,height=0.26\textheight]{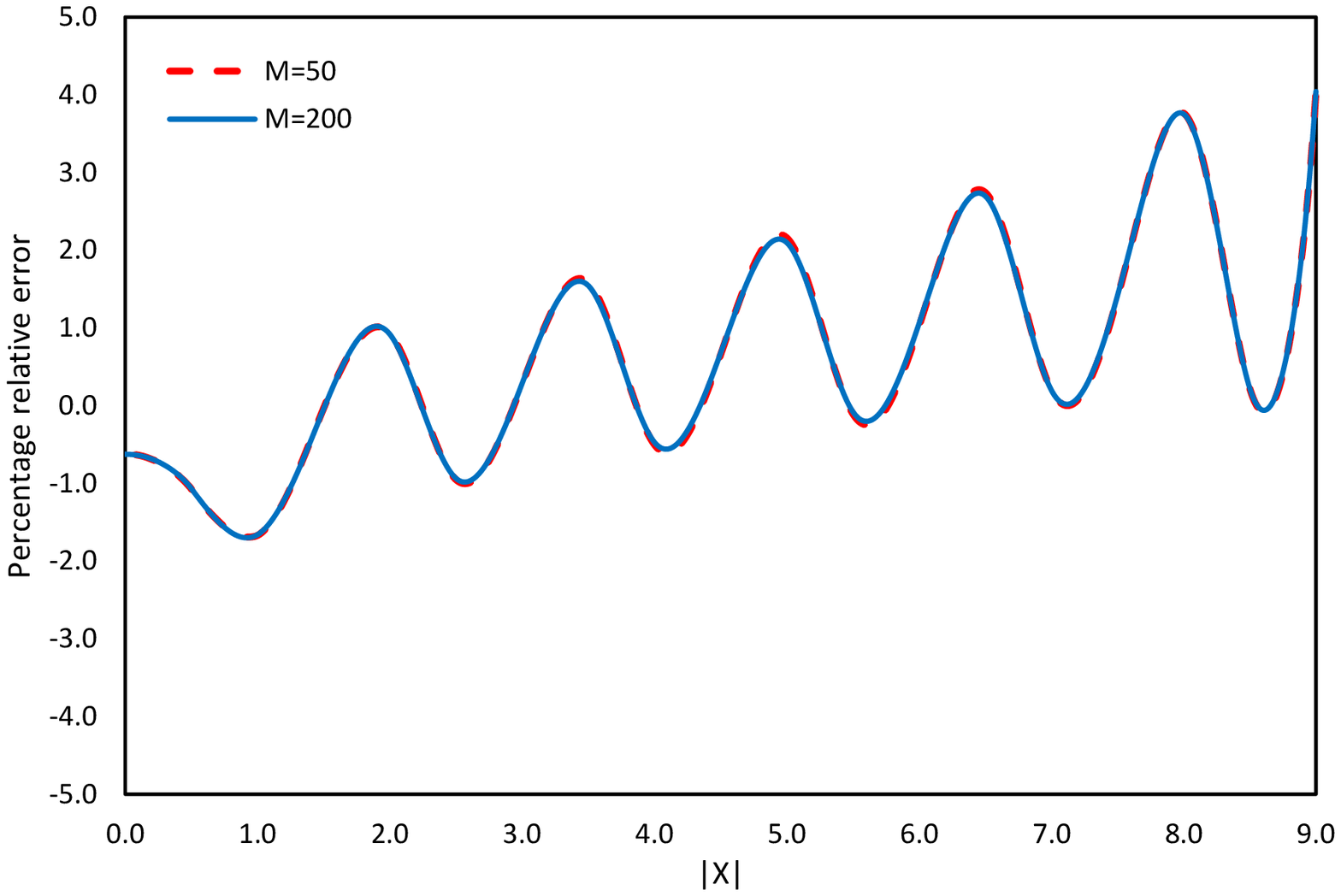}
		\caption{PRE for problem A$_{3}$ with $c = 0.9$.}
	\end{subfigure}
	
	\caption{Percentage relative error (PRE) for problem set A.
}\label{fig4}
\end{figure}

\begin{figure}
	\begin{subfigure}{.48\textwidth}
		\centering
		\includegraphics[width=1\linewidth,height=0.26\textheight]{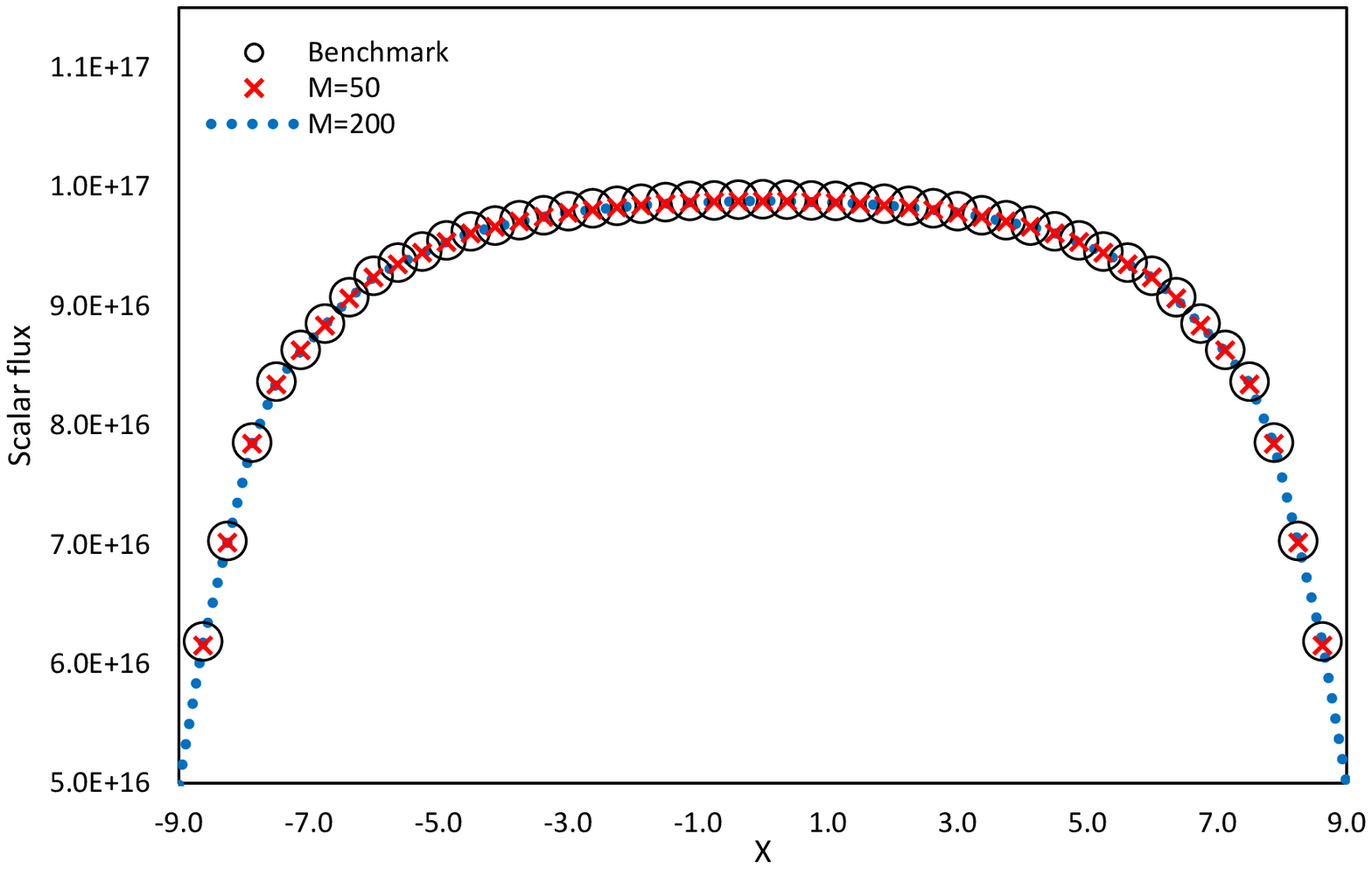}
		\caption{EASF for problem B$_{1}$ with $c = 0.0$.}
	\end{subfigure}
	\begin{subfigure}{.48\textwidth}
		\centering
		\includegraphics[width=1\linewidth,height=0.26\textheight]{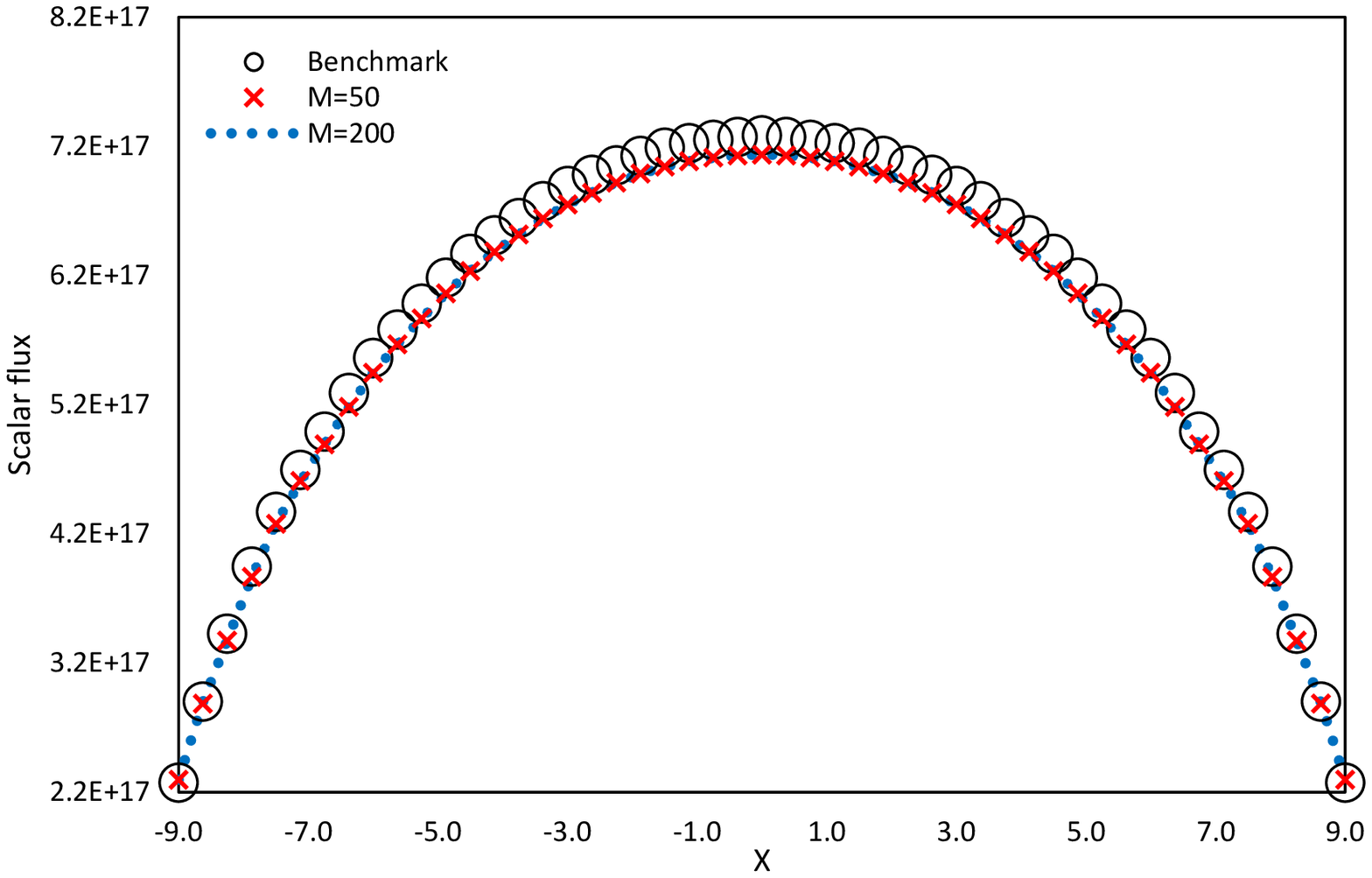}
		\caption{EASF for problem B$_{1}$ with $c = 0.9$.}
	\end{subfigure}
	
	\begin{subfigure}{.48\textwidth}
		\centering
		\includegraphics[width=1\linewidth,height=0.26\textheight]{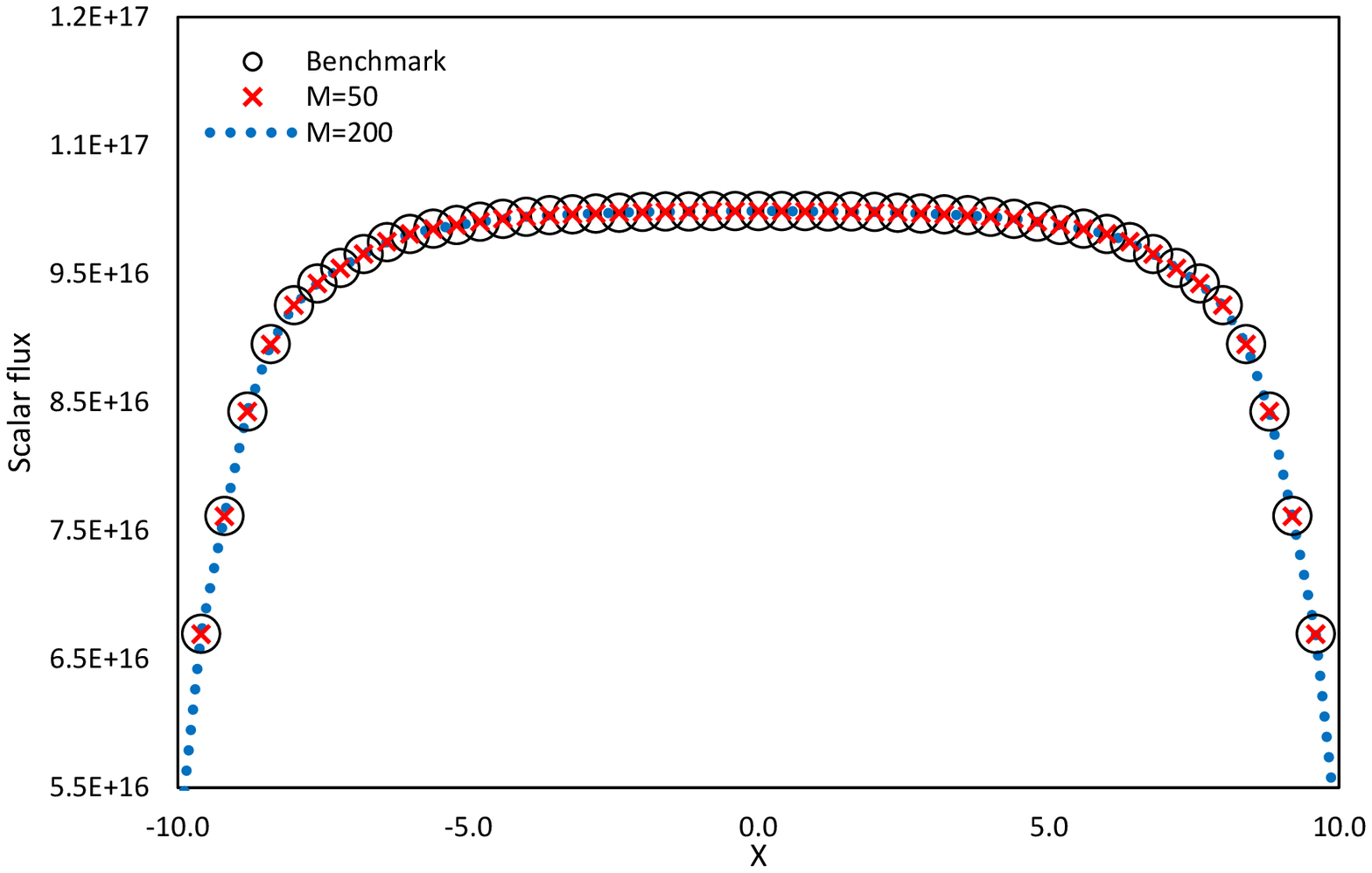}
		\caption{EASF for problem B$_{2}$ with $c = 0.0$.}
	\end{subfigure}
	\begin{subfigure}{.48\textwidth}
		\centering
		\includegraphics[width=1\linewidth,height=0.26\textheight]{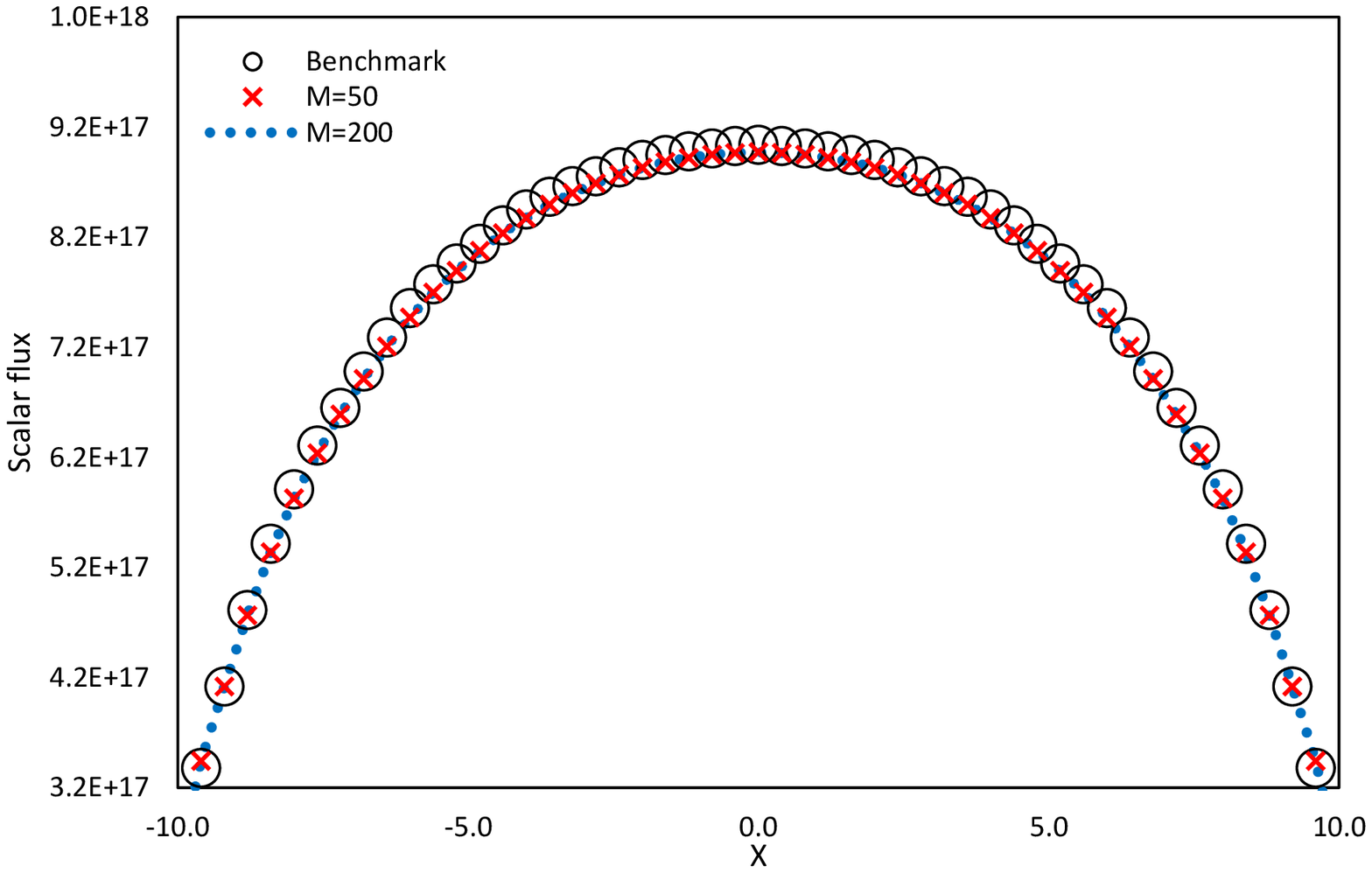}
		\caption{EASF for problem B$_{2}$ with $c = 0.9$.}
	\end{subfigure}
	
	\begin{subfigure}{.48\textwidth}
		\centering
		\includegraphics[width=1\linewidth,height=0.26\textheight]{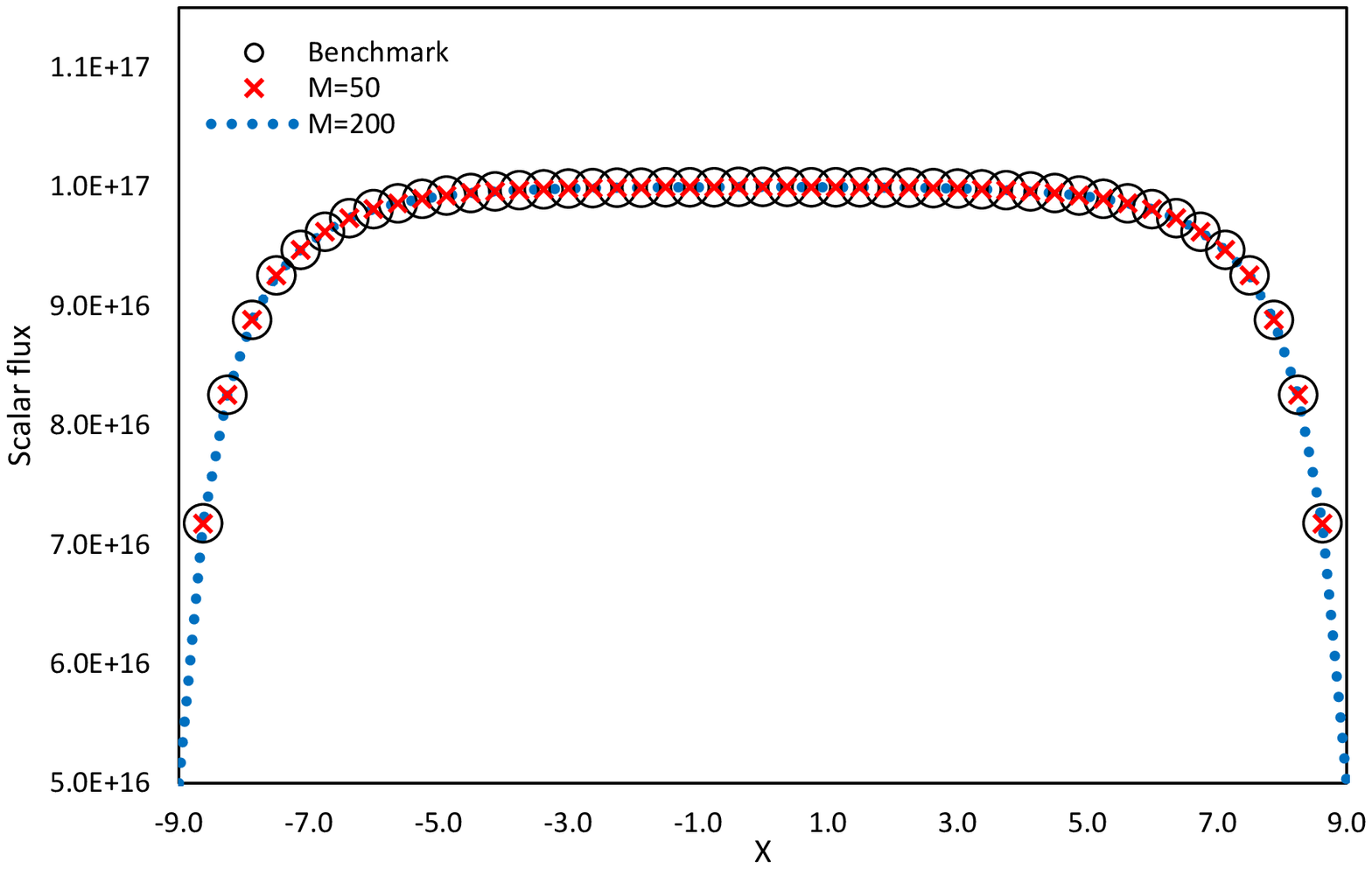}
		\caption{EASF for problem B$_{3}$ with $c = 0.0$.}
	\end{subfigure}
	\begin{subfigure}{.48\textwidth}
		\centering
		\includegraphics[width=1\linewidth,height=0.26\textheight]{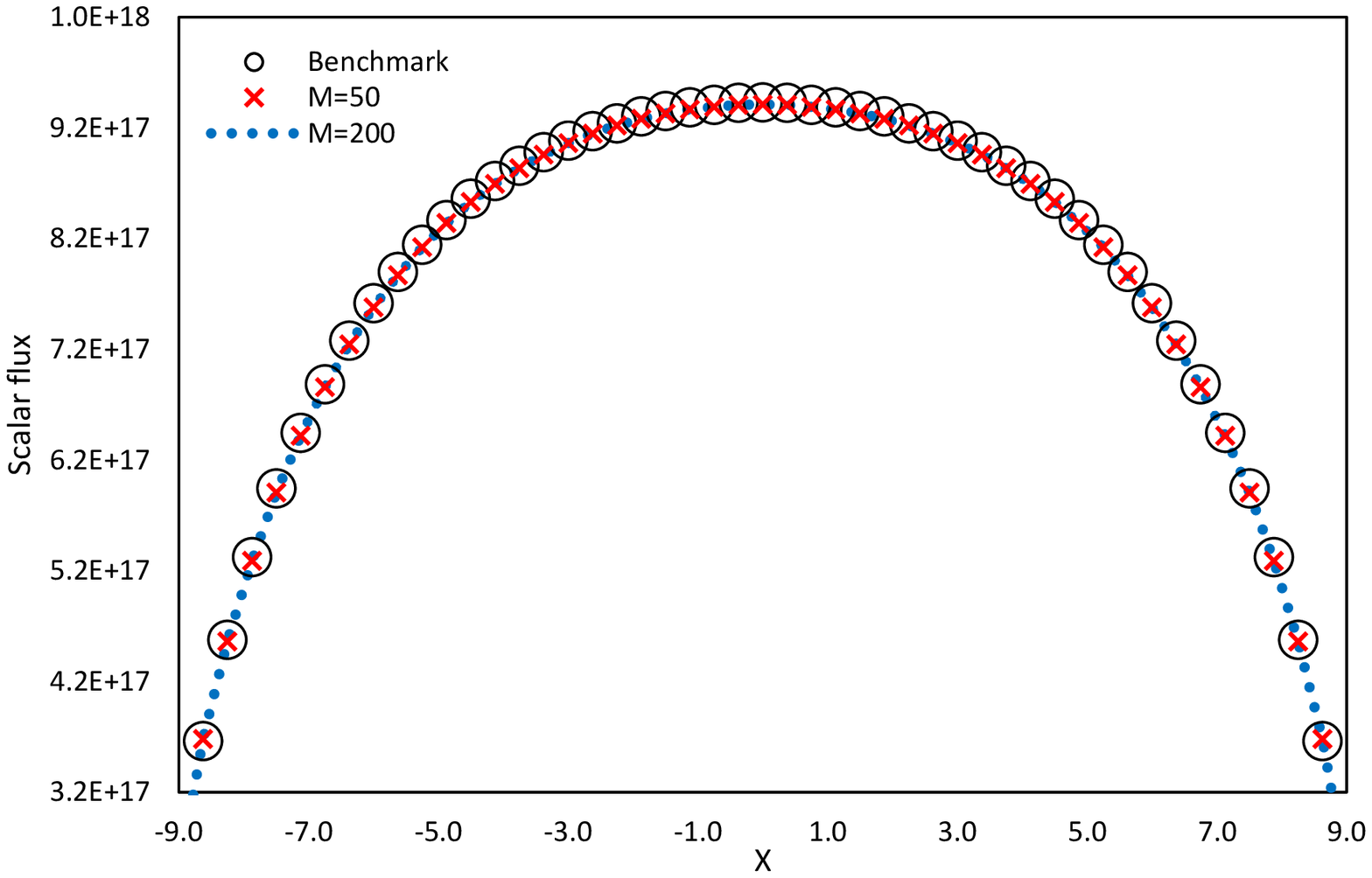}
		\caption{EASF for problem B$_{3}$ with $c = 0.9$.}
	\end{subfigure}
	
	\caption{Ensemble-averaged scalar fluxes (EASF) for problem set B.
}\label{fig5}
\end{figure}

\begin{figure}
	\begin{subfigure}{.48\textwidth}
		\centering
		\includegraphics[width=1\linewidth,height=0.26\textheight]{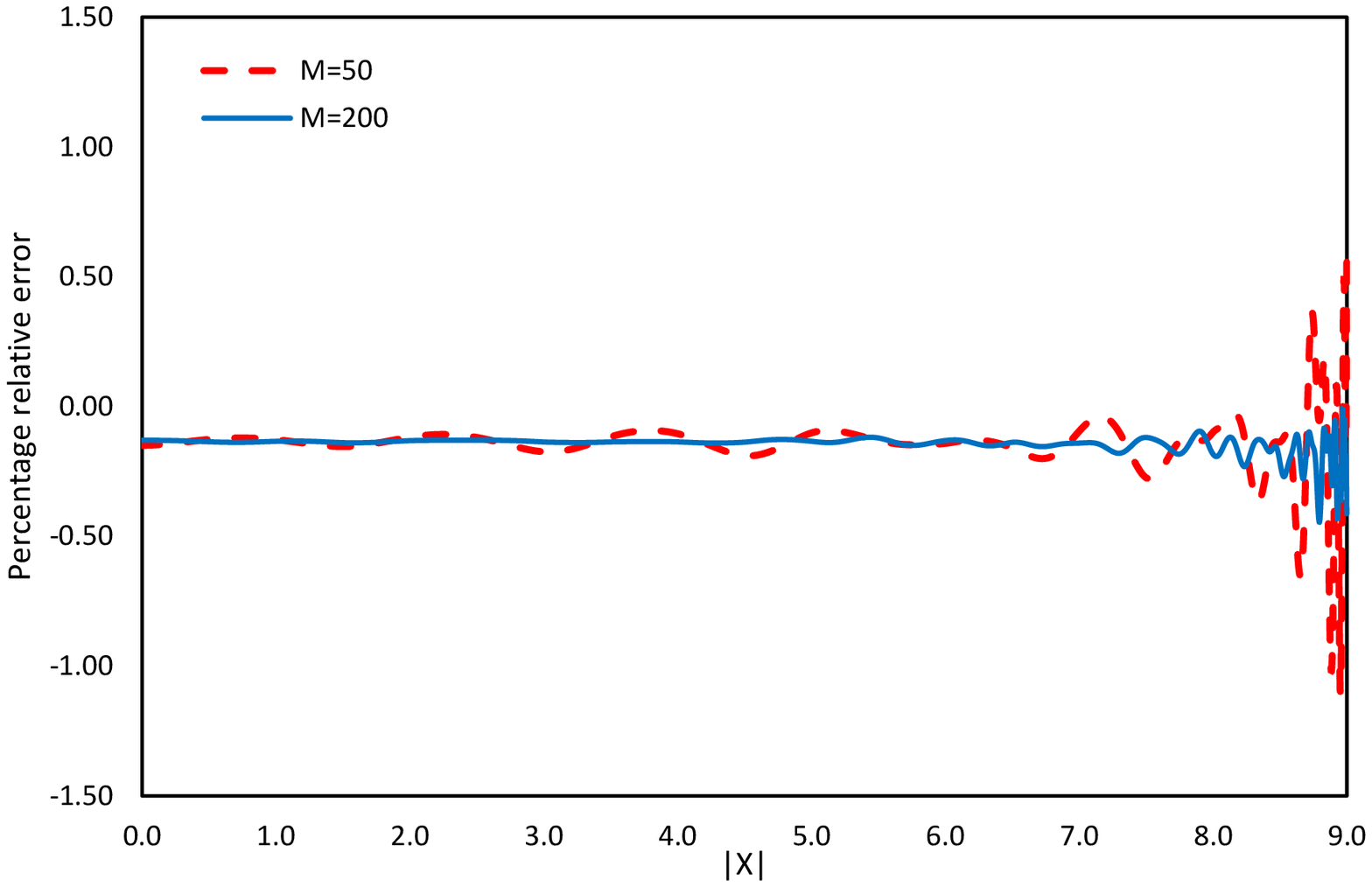}
		\caption{PRE for problem B$_{1}$ with $c = 0.0$.}
	\end{subfigure}
	\begin{subfigure}{.48\textwidth}
		\centering
		\includegraphics[width=1\linewidth,height=0.26\textheight]{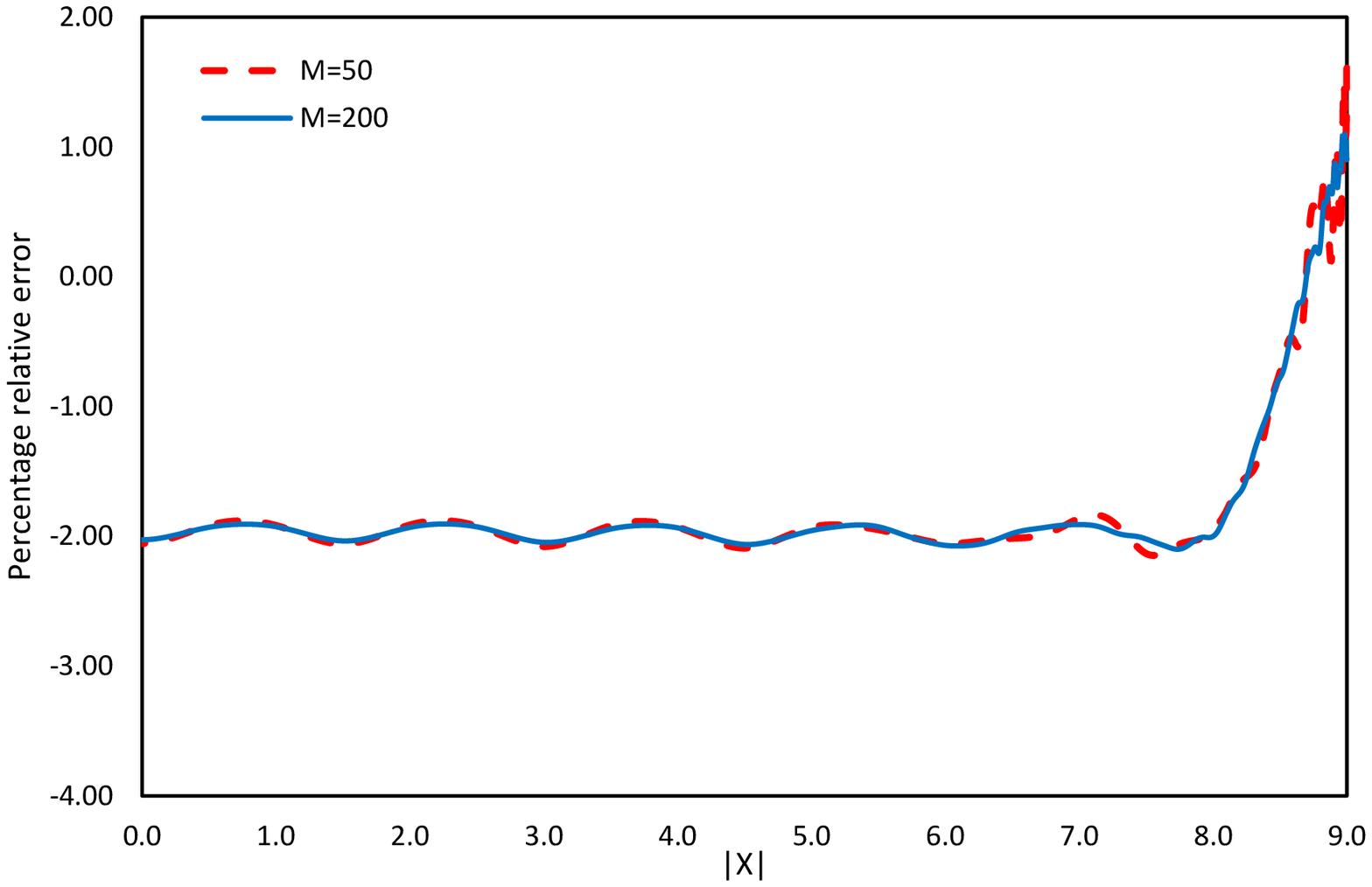}
		\caption{PRE for problem B$_{1}$ with $c = 0.9$.}
	\end{subfigure}
	
	\begin{subfigure}{.48\textwidth}
		\centering
		\includegraphics[width=1\linewidth,height=0.26\textheight]{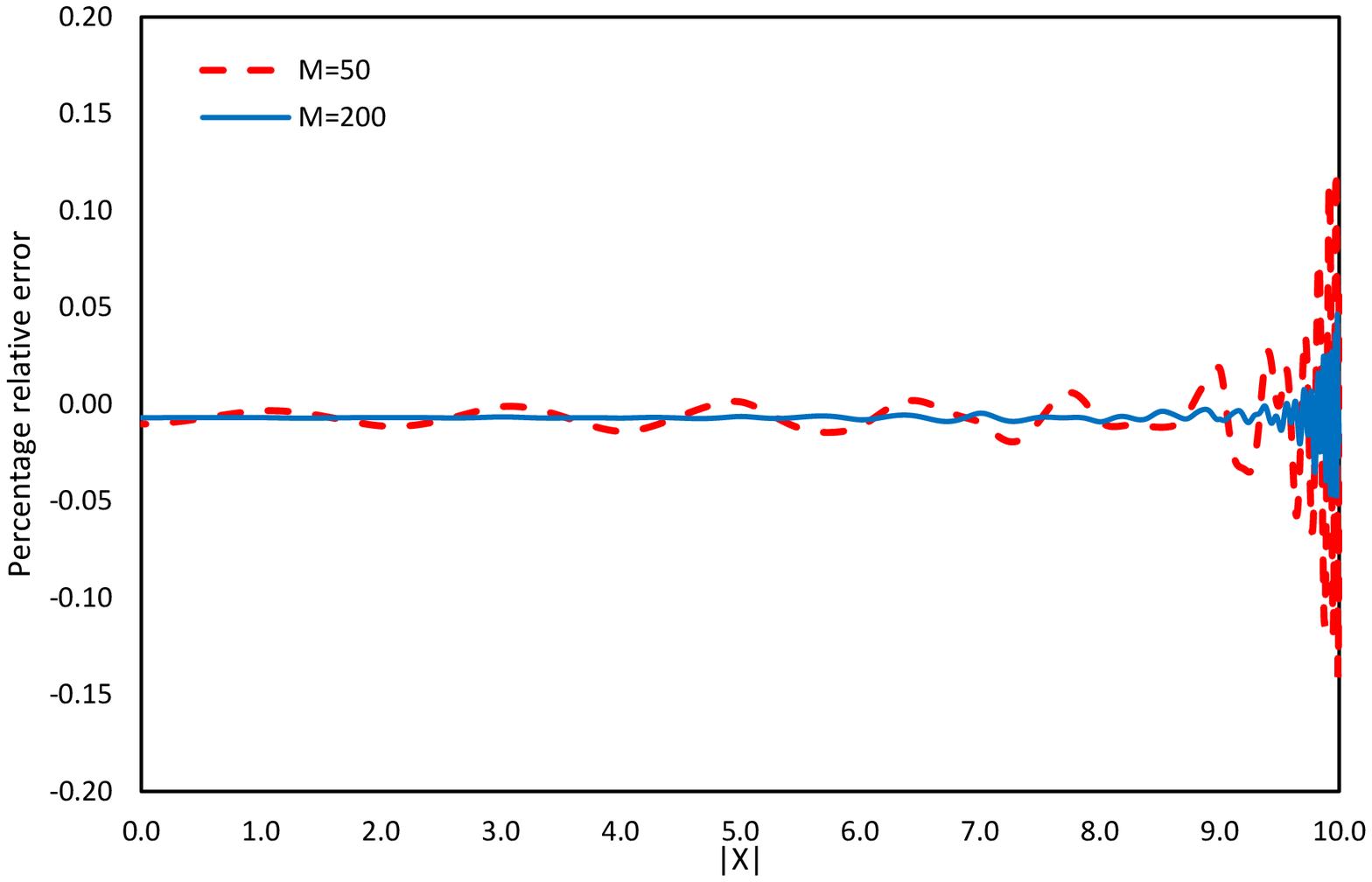}
		\caption{PRE for problem B$_{2}$ with $c = 0.0$.}
	\end{subfigure}
	\begin{subfigure}{.48\textwidth}
		\centering
		\includegraphics[width=1\linewidth,height=0.26\textheight]{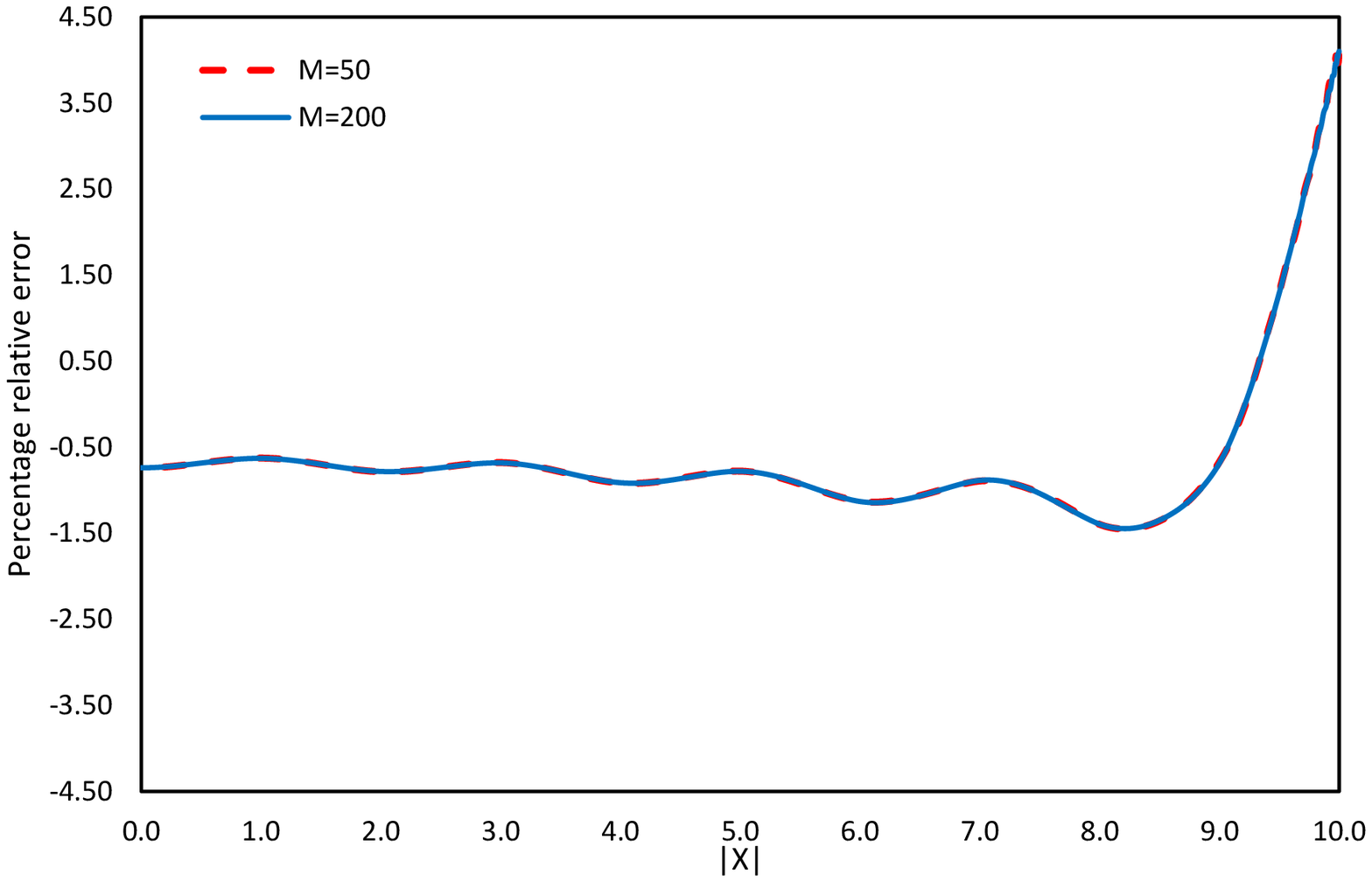}
		\caption{PRE for problem B$_{2}$ with $c = 0.9$.}
	\end{subfigure}
	
	\begin{subfigure}{.48\textwidth}
		\centering
		\includegraphics[width=1\linewidth,height=0.26\textheight]{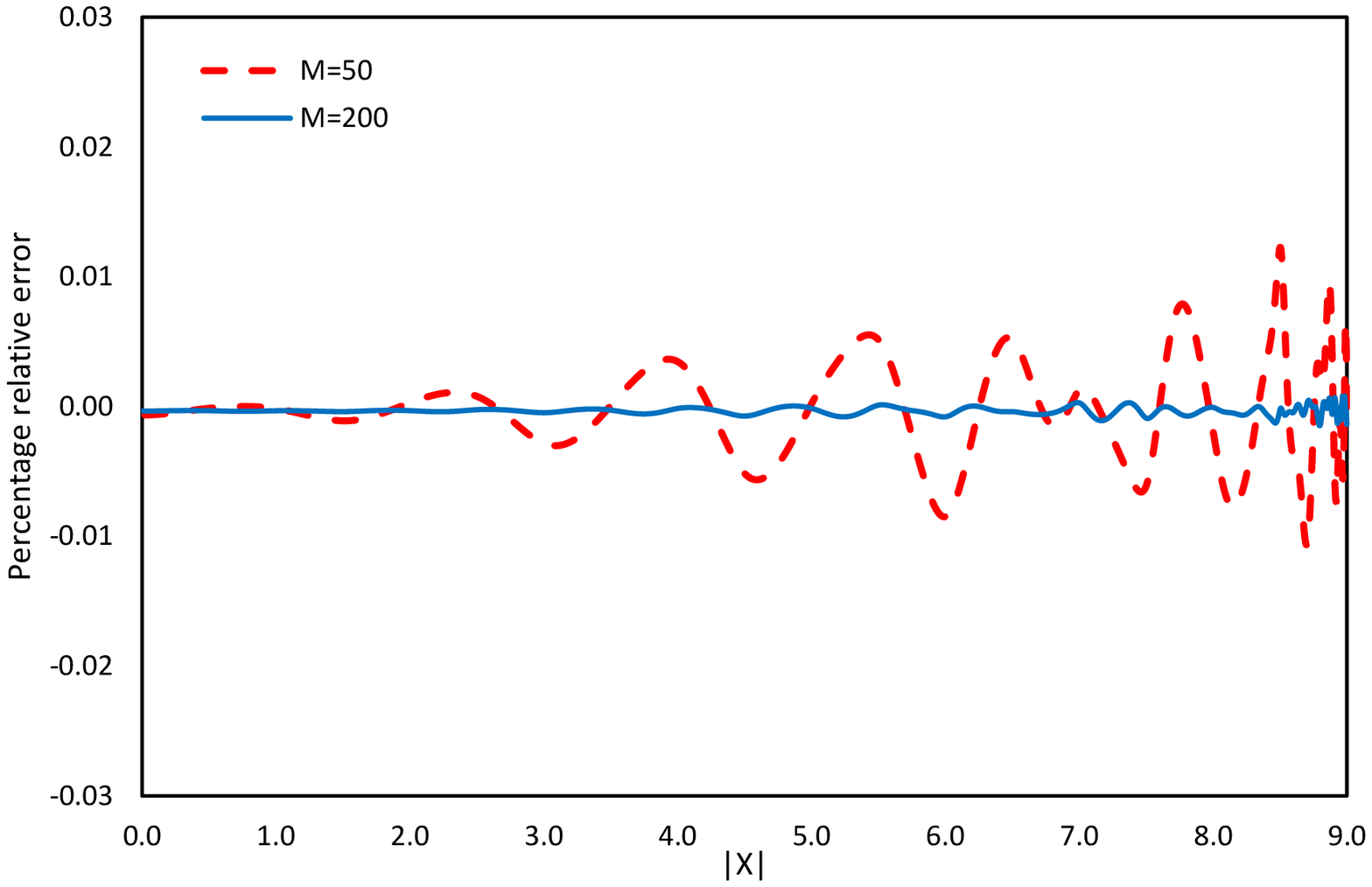}
		\caption{PRE for problem B$_{3}$ with $c = 0.0$.}
	\end{subfigure}
	\begin{subfigure}{.48\textwidth}
		\centering
		\includegraphics[width=1\linewidth,height=0.26\textheight]{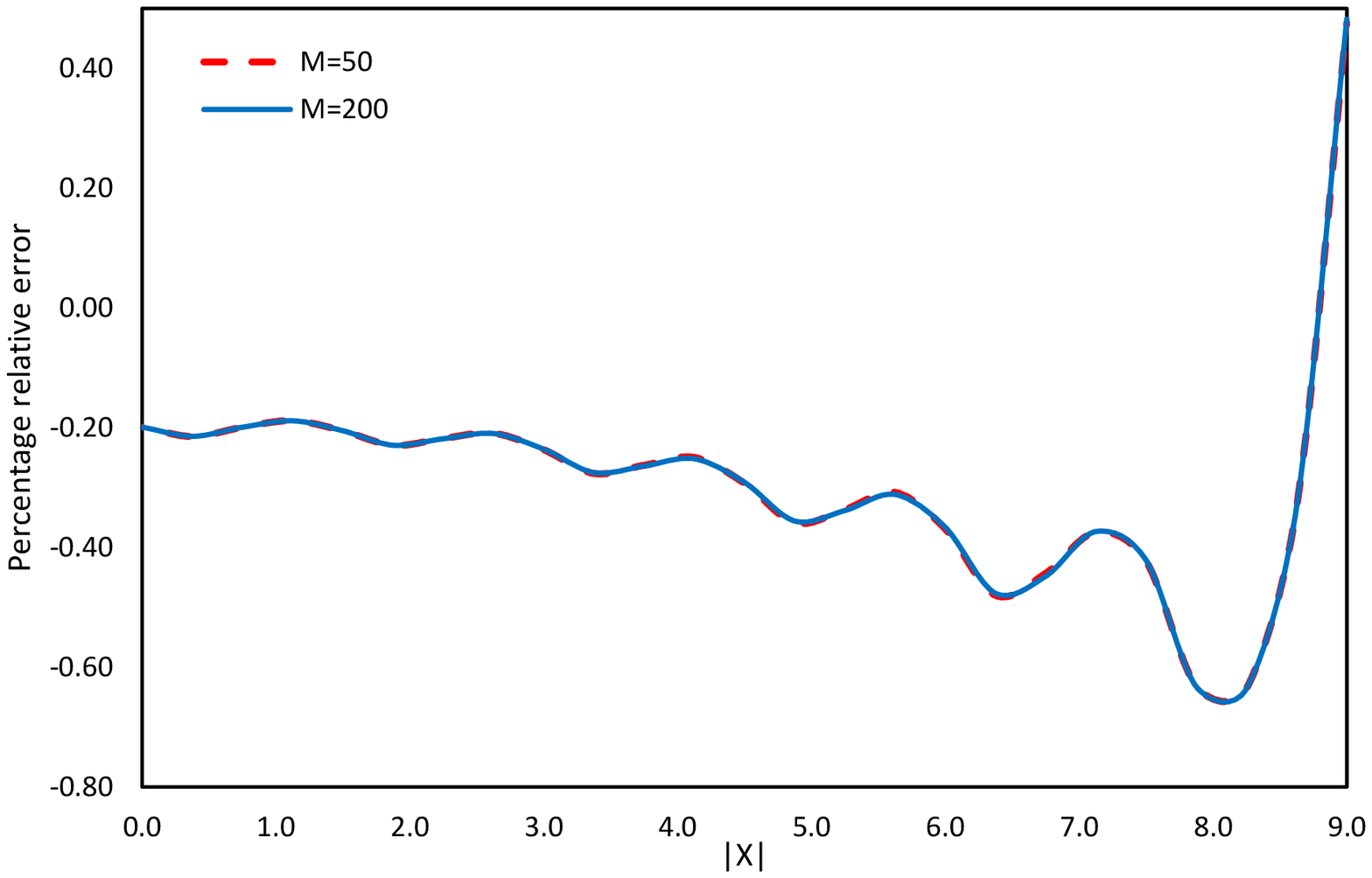}
		\caption{PRE for problem B$_{3}$ with $c = 0.9$.}
	\end{subfigure}
	
	\caption{Percentage relative error (PRE) for problem set B.
}\label{fig6}
\end{figure}

\Cref{fig3} depicts the ensemble-averaged scalar fluxes obtained for problem set A, with \cref{fig4} showing the percentage relative error of the nonclassical approach with respect to the benchmark solutions.
Similarly, \cref{fig5} illustrates the ensemble-averaged scalar fluxes for problem set B, with corresponding percentage relative errors presented in \cref{fig6}. 
Explicit values of these results are given in \cref{tab3,tab4} for easy comparison. 

The best accuracy in both sets of problems is obtained in purely absorbing systems, with $c = 0$.
In these cases, increasing the value of $M$ from $50$ to $200$ shows a clear improvement in the accuracy of the solution.
This can be seen in the left-hand plots in \cref{fig4,fig6} and from the values in \cref{tab3,tab4}.

For this class of test problems, it has been shown \cite{vaskry_17} that the accuracy of the nonclassical model will deteriorate as the system becomes more diffusive, underestimating the maximum value located at $x=0$.
This is confirmed by the numerical results; the estimates obtained for the cases with $c=0.9$ are more inaccurate than the ones generated for purely absorbing cases.
Moreover, there is virtually no improvement in the accuracy of the diffusive solutions upon increasing $M$. 

Finally, as $|x|$ increases, one can see a clear difference in terms of accuracy between solutions of problem sets A and B.
The sinuous shape seen in the relative errors is a consequence of the  periodic structure of the random systems; however, the amplitude of the errors is much larger in problem set A.
This is due to the neutron sources in problem set B being inserted upon all spatial domain, which has two direct effects in the solution: (i) it increases the contribution of unscattered neutrons in the scalar flux, smoothing the error; and (ii) it decreases the propagation of boundary effects. 

\section{Discussion}\label{sec5}
\setcounter{section}{5}
\setcounter{equation}{0} 

In this paper, we have introduced a spectral approach that allows us to numerically solve the nonclassical transport equation using traditional (classical)   methods.
By representing the nonclassical flux as a series of Laguerre polynomials in the free-path variable $s$, we obtain a nonclassical equation that has the form of a classical transport equation.
We describe a numerical methodology to solve this equation using a discrete ordinates ($S_N$) formulation in combination with the diamond difference method and a source iteration scheme.
This was used to solve both classical and nonclassical transport problems in slab geometry, thus proving to be a good tool and validating the proposed approach.
To our knowledge, this is the first time deterministic numerical results have been given for the nonclassical transport equation (\ref{1.1}) in slab geometry.

It is important to note that the goal of this paper is \textit{not} to investigate the accuracy of the nonclassical transport equation; this has been partly addressed elsewhere \cite{vaskry_17} and will be the subject of future work.
Here, we are concerned with the development of an efficient approach to solve the nonclassical transport equation in a deterministic fashion--this work is a first step in that direction.
We remark that the choice of using discrete ordinates or diamond differences is not binding; one can choose several different numerical methodologies to solve the nonclassical equations introduced here.

Modifications of the spectral approach presented in \cref{sec2} are possible and may yield interesting results.
For instance, we can modify \cref{2.2} to define
\bsub
\bal
&\tau(\uom,s) \equiv \alpha + \Sigma_t(\uom,s)\,,
\nal
where $\alpha$ is a constant, and
\bal
&\Psi(\ux,\uom,s)\equiv \hat\psi(\ux,\uom,s)e^{-\int_0^s \tau(\uom,s')ds'}.
\nal 
\nsub
If we expand $\hat\psi$ by a truncated series of Laguerre polynomials, 
\bal
\hat\psi(\ux,\uom,s) = \sum_{m=0}^{M} \hat\psi_m(\ux,\uom)L_m(s)\,,
\nal
and perform the same steps as described in \cref{sec2}, we obtain
\bsub\label[pluraleq]{5.3}
\bal
&\uom\cdot\unab\hat\psi_m(\ux,\uom) + (1-\alpha)\hat\psi_m(\ux,\uom) = \hat S(\ux,\uom) + \f{Q(\ux)}{4\pi} - \sum_{j=0}^{m-1}\hat\psi_j(\ux,\uom), \,\,\, m=0,1, ...,M\,,\label{5.3a}\\
&\hat\psi_m(\ux,\uom) =  \Psi^{b}(\ux,\uom),\quad \ux\in\partial V,\; \un \cdot \uom <0,\; m=0,1, ...,M \,\label{5.3b},
\nal
where
\bal\label{5.3c}
\hat S(\ux,\uom) =  c\int_{4\pi}\int_{0}^{\infty}P(\uom'\cdot\uom)p(\uom',s')e^{-\alpha s'}\textstyle{\sum_{k=0}^{M}\limits}\hat\psi_{k}(\ux,\uom')L_{k}(s')ds'd\uom'\,.
\nal
\nsub
The classical angular flux is given by
\bal\label{5.4}
\Psi_{c}(\ux,\uom) = \int_{0}^{\infty}\Psi(\ux,\uom,s)ds = \int_{0}^{\infty}e^{-\int_{0}^{s}\tau(\uom,s')ds'}\sum_{m=0}^{M}\hat\psi_{m}(\ux,\uom)L_{m}(s)ds\,.
\nal
Assuming classical transport ($\Sigma_t$ independent of $s$ and $\uom$), we can choose $\alpha=1-\Sigma_t$ and $M=0$ to rewrite \cref{5.3} as
\bsub\label[pluraleq]{5.5}
\bal
&\uom\cdot\unab\hat\psi_0(\ux,\uom) + \Sigma_t\hat\psi_0(\ux,\uom) = c\int_{4\pi}P(\uom'\cdot\uom)\Sigma_t\hat\psi_{0}(\ux,\uom')d\uom' + \f{Q(\ux)}{4\pi}\,,\\
&\hat\psi_0(\ux,\uom) =  \Psi^{b}(\ux,\uom),\quad \ux\in\partial V,\; \un \cdot \uom <0\,.
\nal
\nsub
In this case, \cref{5.4} yields $\Psi_c(\ux,\uom) = \hat\psi_0(\ux,\uom)$; therefore, \cref{5.5} represent the classical transport equation for the classical flux as given by \cref{1.3}.
This demonstrates that, for classical transport problems, a proper choice of $\alpha$ and $M$ in \cref{5.3} shall yield the correct solution without truncation errors in the variable $s$.

We remark that, in this paper, we have validated the proposed approach for problems in which the free-path distribution $p(\uom, s)$ is either an exponential representing classical transport (\cref{sec41}) or a nonexponential that decays in an exponential fashion as $s\rightarrow\infty$ (\cref{sec42}).
In these cases, the improper integral in \cref{rev2.1} can be written as a linear combination of the raw moments of $p(\uom, s)$.
This choice of nonclassical problem for validation was made for two reasons: (i) it represents a transport problem whose benchmark solution can be obtained deterministically, and (ii) the free-path distribution is given analitically by \cref{4.1}.

Nevertheless, there are several important applications \cite{davxu_14,xudav_16} in which the free-path distribution has a large tail, decaying algebraically as $s\rightarrow \infty$:
\bal
p(\uom,s) \geq \f{\text{constant}}{s^{k+1}} \quad \text{for } s\gg 1\,.
\nal
In these cases, the moments of order $k$ and larger will not exist.
Therefore, the approximation
\bal
\int_0^\infty p(\uom,s)\psi(\ux,\uom,s)ds \approx \sum_{k=0}^M\psi_k(\ux,\uom)\int_0^L p(\uom,s)L_k(s)ds\,,
\nal
where $L$ is a constant larger than the maximum chord length of the system, may compromise the efficiency of the method. 
We acknowledge this possibility and intend to perform a deeper investigation of the proposed method's capabilities in addressing such problems. 
However, since benchmark solutions for this type of transport problem would need to be obtained through a Monte Carlo calculation, this task is beyond the scope of this purely deterministic paper and shall be pursued in future work.

Further work will also need to be done to investigate how well this approach performs in multi-dimensional nonclassical systems. 
In order to address that, the next steps of this work include: (i) implementing an acceleration scheme to improve the efficiency of the method; (ii) investigating the use of coarse-mesh methods, such as Response Matrix \cite{mansur}; and (iii) performing a full convergence analysis of the method. 
We also intend to drop the periodic assumption and investigate results in more realistic random media; however, this will require a sophisticated numerical approach to estimate the free-path distribution $p(\uom,s)$.

\section*{Acknowledgements}

R.~Vasques acknowledges support under award number NRC-HQ-84-15-G-0024 from the Nuclear Regulatory Commission.
R.~N.~Slaybaugh acknowledges support under award number
NRC-HQ-84-14-G-0052 from the Nuclear Regulatory Commission.
This study was financed in part by the Coordena\c{c}\~ao de
Aperfei\c{c}oamento de Pessoal de N\'ivel Superior - Brasil (CAPES) -
Finance Code 001.
L.~R.~C.~Moraes and R.~C.~Barros also would like to express their gratitude to the support of Conselho Nacional de Desenvolvimento Cient\'ifico e Tecnol\'ogico - Brasil (CNPq) and Funda\c{c}\~ao Carlos Chagas Filho de Amparo \`a Pesquisa do Estado do Rio de Janeiro - Brasil (FAPERJ).

\newpage


\bibliography{JCP_bib}

\newpage

\newpage
\begin{landscape}
\begin{table}[htbp]
	\centering
	\caption{Classical scalar fluxes for the slab-geometry classical transport problem. }
	\scalebox{0.85}{

      \begin{tabular}{|c|c|c|c|c|c|c|c|}
      	\toprule
      	$|\text{x}|$   & Classical transport equation & \multicolumn{3}{c|}{Nonclassical transport equation} & \multicolumn{3}{c|}{Relative error } \\
      	(cm)  &  (neutrons/cm$^{2}$s) & \multicolumn{3}{c|}{(neutrons/cm$^{2}$s)} & \multicolumn{3}{c|}{(\%) } \\
      	\cmidrule{3-8}          &       & $M$=0   & $M$=50  & $M$=100 & $M$=0   & $M$=50  & $M$=100 \\
      	\midrule
      	\multicolumn{8}{|c|}{$c$ = 0.0} \\
      	\midrule
      	0.0   & 6.734502E+16 & 6.734502E+16 & 6.734502E+16 & 6.734502E+16 & -2.061153E-07 & 8.305695E-07 & -2.316133E-09 \\
    	\midrule
    	1.0     & 1.2672661E+16 & 1.2672661E+16 & 1.2672661E+16 & 1.2672661E+16 & -2.0611535E-07 & -6.5724395E-07 & -1.4390150E-06 \\
      	\midrule
      	2.0   & 2.664948E+15 & 2.664948E+15 & 2.664948E+15 & 2.664948E+15 & -2.061155E-07 & -2.378777E-06 & 2.347877E-06 \\
      	\midrule
      	3.0   & 6.997904E+14 & 6.997904E+14 & 6.997902E+14 & 6.997905E+14 & -2.061153E-07 & -2.423450E-05 & 6.092642E-06 \\
      	\midrule
      	4.0   & 2.011571E+14 & 2.011571E+14 & 2.011571E+14 & 2.011571E+14 & -2.061156E-07 & 2.127259E-05 & 3.275557E-05 \\
      	\midrule
      	5.0   & 6.084981E+13 & 6.084981E+13 & 6.084979E+13 & 6.084978E+13 & -2.061152E-07 & -4.467129E-05 & -5.366992E-05 \\
      	\midrule
      	6.0   & 1.902634E+13 & 1.902634E+13 & 1.902632E+13 & 1.902633E+13 & -2.061153E-07 & -8.692087E-05 & -5.532571E-05 \\
      	\midrule
      	7.0   & 6.089504E+12 & 6.089504E+12 & 6.089505E+12 & 6.089537E+12 & -2.061152E-07 & 1.857005E-05 & 5.378618E-04 \\
      	\midrule
      	8.0   & 1.983150E+12 & 1.983150E+12 & 1.983146E+12 & 1.983127E+12 & -2.061151E-07 & -1.905777E-04 & -1.142114E-03 \\
      	\midrule
      	9.0   & 6.546034E+11 & 6.546034E+11 & 6.545958E+11 & 6.546020E+11 & -2.061152E-07 & -1.166810E-03 & -2.140537E-04 \\
      	\midrule
      	10.0  & 2.184099E+11 & 2.184099E+11 & 2.184053E+11 & 2.184060E+11 & -2.061152E-07 & -2.118095E-03 & -1.786618E-03 \\
      	\midrule
      	\multicolumn{8}{|c|}{$c = 0.9$} \\
      	\midrule
      	0.0   & 2.719908E+17 & 2.719907E+17 & 2.719908E+17 & 2.719908E+17 & -1.916088E-05 & 2.227138E-06 & 1.924450E-06 \\
      	\midrule
      	1.0   & 1.537115E+17 & 1.537114E+17 & 1.537115E+17 & 1.537115E+17 & -3.315645E-05 & 3.321180E-06 & 3.264306E-06 \\
      	\midrule
      	2.0   & 8.680613E+16 & 8.680608E+16 & 8.680613E+16 & 8.680613E+16 & -5.609675E-05 & 5.742565E-06 & 5.833409E-06 \\
      	\midrule
      	3.0   & 5.065063E+16 & 5.065058E+16 & 5.065063E+16 & 5.065064E+16 & -8.992331E-05 & 8.735536E-06 & 9.487710E-06 \\
      	\midrule
      	4.0   & 2.978408E+16 & 2.978404E+16 & 2.978408E+16 & 2.978408E+16 & -1.396228E-04 & 1.485754E-05 & 1.487591E-05 \\
      	\midrule
      	5.0   & 1.754675E+16 & 1.754672E+16 & 1.754676E+16 & 1.754676E+16 & -2.101284E-04 & 2.128752E-05 & 2.129112E-05 \\
      	\midrule
      	6.0   & 1.032058E+16 & 1.032055E+16 & 1.032059E+16 & 1.032059E+16 & -3.049518E-04 & 3.112824E-05 & 3.108565E-05 \\
      	\midrule
      	7.0   & 6.024440E+15 & 6.024414E+15 & 6.024442E+15 & 6.024443E+15 & -4.229096E-04 & 4.217371E-05 & 4.354501E-05 \\
      	\midrule
      	8.0   & 3.433063E+15 & 3.433044E+15 & 3.433065E+15 & 3.433065E+15 & -5.529738E-04 & 5.372110E-05 & 5.202548E-05 \\
      	\midrule
      	9.0   & 1.805655E+15 & 1.805643E+15 & 1.805656E+15 & 1.805656E+15 & -6.683423E-04 & 6.025153E-05 & 6.203616E-05 \\
      	\midrule
      	10.0  & 5.845608E+14 & 5.845566E+14 & 5.845611E+14 & 5.845611E+14 & -7.046261E-04 & 5.564129E-05 & 5.684071E-05 \\
      	\bottomrule
      \end{tabular}%
}
	\label{tab1}%
\end{table}%
\end{landscape}

\newpage

\newpage

\begin{table}[htbp]
	\centering
	\caption{Parameters of nonclassical test problems.}
		\scalebox{1}{
    \begin{tabular}{|c|c|c|c|c|c|c|}
	\toprule
	\multicolumn{2}{|c|}{Problem} & Space domain limits & $\ell_{1}$    & $\ell_{2}$    & $x_{1}$    & $x_{2}$ \\
	\multicolumn{1}{|c}{} &       & (cm)  & (cm)  & (cm)  & (cm)  & (cm) \\
	\midrule
	\multirow{3}[6]{*}{A} & A$_{1}$    & [-9.0, 9.0] & 0.5   & 1.0   & -0.5  & 0.5 \\
	\cmidrule{2-7}          & A$_{2}$     & [-10.0, 10.0] & 1.0   & 1.0   & -0.5  & 0.5 \\
	\cmidrule{2-7}          & A$_{3}$     & [-9.0, 9.0] & 1.0   & 0.5   & -0.5  & 0.5 \\
	\midrule
	\multirow{3}[6]{*}{B} & B$_{1}$     & [-9.0, 9.0] & 0.5   & 1.0   & -9.0  & 9.0 \\
	\cmidrule{2-7}          & B$_{2}$     & [-10.0, 10.0] & 1.0   & 1.0   & -10.0 & 10.0 \\
	\cmidrule{2-7}          & B$_{3}$     & [-9.0, 9.0] & 1.0   & 0.5   & -9.0  & 9.0 \\
	\bottomrule
\end{tabular}%
\label{tab2}%
}
\end{table}%

\newpage

\newpage

\newpage

\newpage

\newpage

\begin{landscape}
	\begin{table}[htbp]
		\centering
		\caption{Ensemble-Averaged scalar fluxes for Problem Set A.}\label{tab3}
		\scalebox{0.78}{
    \begin{tabular}{|c|c|c|c|c|c|c|}
	\toprule
	\multirow{3}[4]{*}{Problem} & $|\text{x}|$    & \multicolumn{1}{c|}{Benchmark} & \multicolumn{2}{c|}{Nonclassical transport equation} & \multicolumn{2}{c|}{Relative error } \\
	& (cm)  & \multicolumn{1}{c|}{(neutrons/cm$^{2}$s)} & \multicolumn{2}{c|}{(neutrons/cm$^{2}$s)} & \multicolumn{2}{c|}{(\%)} \\
	\cmidrule{4-7}          &       &       & \multicolumn{1}{c|}{$M$=50} & \multicolumn{1}{c|}{$M$=200} & $M$=50  & $M$=200 \\
	\midrule
	\multirow{8}[16]{*}{A$_{1}$} & \multicolumn{6}{c|}{$c$ = 0.0} \\
	\cmidrule{2-7}          & 0.0   & 2.941938E+16 & 2.948194E+16 & 2.936095E+16 & 2.126596E-01 & -1.985964E-01 \\
	\cmidrule{2-7}          & 4.5   & 1.756486E+15 & 1.743672E+15 & 1.745583E+15 & -7.295201E-01 & -6.207207E-01 \\
	\cmidrule{2-7}          & 9.0  & 2.252977E+14 & 2.187016E+14 & 2.198255E+14 & -2.927735E+00 & -2.428882E+00 \\
	\cmidrule{2-7}          & \multicolumn{6}{c|}{$c$ = 0.9} \\
	\cmidrule{2-7}          & 0.0   & 9.745114E+16 & 9.595979E+16 & 9.586788E+16 & -1.530351E+00 & -1.624665E+00 \\
	\cmidrule{2-7}          & 4.5   & 3.389651E+16 & 3.322013E+16 & 3.327105E+16 & -1.995425E+00 & -1.845201E+00 \\
	\cmidrule{2-7}          & 9.0  & 7.610961E+15 & 7.650189E+15 & 7.647939E+15 & 5.154120E-01 & 4.858457E-01 \\
	\midrule
	\multirow{8}[16]{*}{A$_{2}$} & \multicolumn{6}{c|}{$c$ = 0.0} \\
	\cmidrule{2-7}          & 0.0   & 3.890079E+16 & 3.891017E+16 & 3.889305E+16 & 2.411346E-02 & -1.988391E-02 \\
	\cmidrule{2-7}          & 5.0   & 6.598115E+14 & 6.660539E+14 & 6.591396E+14 & 9.460837E-01 & -1.018436E-01 \\
	\cmidrule{2-7}          & 10.0  & 2.999265E+13 & 3.028603E+13 & 2.994294E+13 & 9.781649E-01 & -1.657549E-01 \\
	\cmidrule{2-7}          & \multicolumn{6}{c|}{$c$ = 0.9} \\
	\cmidrule{2-7}          & 0.0   & 1.427736E+17 & 1.410333E+17 & 1.410050E+17 & -1.218951E+00 & -1.238745E+00 \\
	\cmidrule{2-7}          & 5.0   & 3.186643E+16 & 3.218562E+16 & 3.214855E+16 & 1.001670E+00 & 8.853233E-01 \\
	\cmidrule{2-7}          & 10.0  & 4.032572E+15 & 4.365284E+15 & 4.367367E+15 & 8.250617E+00 & 8.302263E+00 \\
	\midrule
	\multirow{8}[16]{*}{A$_{3}$} & \multicolumn{6}{c|}{$c$ = 0.0} \\
	\cmidrule{2-7}          & 0.0   & 5.186772E+16 & 5.188694E+16 & 5.186813E+16 & 3.706624E-02 & 7.872449E-04 \\
	\cmidrule{2-7}          & 4.5   & 4.585612E+14 & 4.616722E+14 & 4.585512E+14 & 6.784263E-01 & -2.173120E-03 \\
	\cmidrule{2-7}          & 9.0  & 1.251280E+13 & 1.249927E+13 & 1.248988E+13 & -1.080800E-01 & -1.831402E-01 \\
	\cmidrule{2-7}          & \multicolumn{6}{c|}{$c$ = 0.9} \\
	\cmidrule{2-7}          & 0.0   & 1.908438E+17 & 1.896727E+17 & 1.896511E+17 & -6.136817E-01 & -6.249941E-01 \\
	\cmidrule{2-7}          & 4.5   & 3.302245E+16 & 3.324623E+16 & 3.325332E+16 & 6.776650E-01 & 6.991184E-01 \\
	\cmidrule{2-7}          & 9.0  & 3.180001E+15 & 3.312502E+15 & 3.312802E+15 & 4.166693E+00 & 4.176118E+00 \\
	\bottomrule
\end{tabular}%
	}
	\end{table}%
\end{landscape}

\newpage

\begin{landscape}
	\begin{table}[htbp]
		\centering
		\caption{Ensemble-Averaged scalar fluxes for Problem Set B.}\label{tab4}
		\scalebox{0.78}{
		\begin{tabular}{|c|c|c|c|c|c|c|}
			\toprule
			\multirow{3}[4]{*}{Problem} & $|\text{x}|$    & \multicolumn{1}{c|}{Benchmark} & \multicolumn{2}{c|}{Nonclassical transport equation} & \multicolumn{2}{c|}{Relative error } \\
			& (cm)  & \multicolumn{1}{p{7.285em}|}{(neutrons/cm$^{2}$s)} & \multicolumn{2}{c|}{(neutrons/cm$^{2}$s)} & \multicolumn{2}{c|}{(\%)} \\
			\cmidrule{4-7}          &       &       & \multicolumn{1}{c|}{$M$=50} & \multicolumn{1}{c|}{$M$=200} & $M$=50  & $M$=200 \\
			\midrule
			\multirow{8}[16]{*}{B$_{1}$} & \multicolumn{6}{c|}{$c$ = 0.0} \\
			\cmidrule{2-7}          & 0.0   & 9.893582E+16 & 9.878834E+16 & 9.880716E+16 & -1.490681E-01 & -1.300397E-01 \\
			\cmidrule{2-7}          & 4.5   & 9.625624E+16 & 9.607356E+16 & 9.612430E+16 & -1.897782E-01 & -1.370743E-01 \\
			\cmidrule{2-7}          & 9.0  & 4.998409E+16 & 4.986236E+16 & 4.986238E+16 & -2.435295E-01 & -2.434841E-01 \\
			\cmidrule{2-7}          & \multicolumn{6}{c|}{$c$ = 0.9} \\
			\cmidrule{2-7}          & 0.0   & 7.286388E+17 & 7.136680E+17 & 7.138745E+17 & -2.054628E+00 & -2.026288E+00 \\
			\cmidrule{2-7}          & 4.5   & 6.367297E+17 & 6.234173E+17 & 6.235945E+17 & -2.090757E+00 & -2.062916E+00 \\
			\cmidrule{2-7}          & 9.0  & 2.273773E+17 & 2.296963E+17 & 2.297052E+17 & 1.019883E+00 & 1.023785E+00 \\
			\midrule
			\multirow{8}[16]{*}{B$_{2}$} & \multicolumn{6}{c|}{$c$ = 0.0} \\
			\cmidrule{2-7}          & 0.0   & 9.990035E+16 & 9.989015E+16 & 9.989337E+16 & -1.021726E-02 & -6.988034E-03 \\
			\cmidrule{2-7}          & 5.0   & 9.893674E+16 & 9.893792E+16 & 9.893049E+16 & 1.191124E-03 & -6.325158E-03 \\
			\cmidrule{2-7}          & 10.0  & 4.999981E+16 & 4.999523E+16 & 4.999523E+16 & -9.160875E-03 & -9.156552E-03 \\
			\cmidrule{2-7}          & \multicolumn{6}{c|}{$c$ = 0.9} \\
			\cmidrule{2-7}          & 0.0   & 9.038937E+17
			 & 8.971970E+17   & 8.972188E+17   & -7.408680E-01  & -7.384594E-01  \\
			\cmidrule{2-7}          & 5.0   & 8.050648E+17 & 7.988039E+17
			 & 7.987906E+17	  & -7.776922E-01 & -7.793447E-01 \\
			\cmidrule{2-7}          & 10.0  & 2.386542E+17 & 2.485283E+17
			 & 2.485310E+17  & 4.137392E+00  & 4.138536E+00  \\
			\midrule
			\multirow{8}[16]{*}{B$_{3}$} & \multicolumn{6}{c|}{$c$ = 0.0} \\
			\cmidrule{2-7}          & 0.0   & 9.996817E+16 & 9.996753E+16 & 9.996782E+16 & -6.494109E-04 & -3.513593E-04 \\
			\cmidrule{2-7}          & 4.5   & 9.946734E+16 & 9.946217E+16 & 9.946661E+16 & -5.197731E-03 & -7.356050E-04 \\
			\cmidrule{2-7}          & 9.0  & 4.999998E+16 & 4.999966E+16 & 4.999966E+16 & -6.286861E-04 & -6.282565E-04 \\
			\cmidrule{2-7}          & \multicolumn{6}{c|}{$c$ = 0.9} \\
			\cmidrule{2-7}          & 0.0   & 9.431169E+17 & 9.410914E+17 & 9.410939E+17 & -2.147673E-01 & -2.145026E-01 \\
			\cmidrule{2-7}          & 4.5   & 8.560157E+17 & 8.529432E+17 & 8.529691E+17 & -3.589333E-01 & -3.558992E-01 \\
			\cmidrule{2-7}          & 9.0  & 2.396932E+17 & 2.445678E+17 & 2.445680E+17 & 2.033667E+00 & 2.033756E+00 \\
			\bottomrule
		\end{tabular}%
}
	\end{table}%
\end{landscape}

\end{document}